\documentclass[preprint,12pt,3p]{elsarticle}

\usepackage{amssymb}

\usepackage{amsfonts}
\usepackage{graphicx}
\usepackage{multirow}
\usepackage{bigstrut}
\usepackage{epstopdf}
\usepackage[english]{babel}
\usepackage{combelow}
\usepackage{amsmath}
\usepackage{amsthm}
\usepackage{algorithm}
\usepackage{algorithmicx}
\usepackage[noend]{algpseudocode}
\usepackage{xcolor}

\newdefinition{rmk}{Remark}
\numberwithin{equation}{section}

\journal{J. Comp. Phys.}

\begin{document}

\begin{frontmatter}

\title{Sensitivity-driven adaptive sparse stochastic approximations in plasma microinstability analysis}

\author[label1]{Ionu\cb{t}-Gabriel Farca\cb{s}}
\address[label1]{Department of Informatics, Technical University of Munich, Boltzmannstr. 3, 85748 Garching, Germany}
\ead{farcasi@in.tum.de}


\author[label2]{Tobias G\"orler}
\address[label2]{Max Planck Institute for Plasma Physics, Boltzmannstr. 2, 85748 Garching, Germany}
\ead{tobias.goerler@ipp.mpg.de}

\author[label1]{Hans-Joachim Bungartz}
\ead{bungartz@in.tum.de}

\author[label2,label1,label3]{Frank Jenko}
\address[label3]{Oden Institute for Computational Engineering and Sciences, University of Texas at Austin, 201 E 24th St, Austin, TX 78712, United States}
\ead{frank.jenko@ipp.mpg.de}

\author[label1]{Tobias Neckel}
\ead{neckel@in.tum.de}

\begin{abstract}
Quantifying uncertainty in predictive simulations for real-world problems is of paramount importance - and far from trivial, mainly due to the large number of stochastic parameters and significant computational requirements.
Adaptive sparse grid approximations are an established approach to overcome these challenges.
However, standard adaptivity is based on global information, thus properties such as lower intrinsic stochastic dimensionality or anisotropic coupling of the input directions, which are common in practical applications, are not fully exploited.
We propose a novel structure-exploiting dimension-adaptive sparse grid approximation methodology using Sobol' decompositions in each subspace to introduce a sensitivity scoring system to drive the adaptive process.
By employing sensitivity information, we explore and exploit the anisotropic coupling of the stochastic inputs as well as the lower intrinsic stochastic dimensionality.
The proposed approach is generic, i.e., it can be formulated in terms of arbitrary approximation operators and point sets. 
In particular, we consider sparse grid interpolation and pseudo-spectral projection constructed on (L)-Leja sequences.
The power and usefulness of the proposed method is demonstrated by applying it to the analysis of gyrokinetic microinstabilities in fusion plasmas, one of the key scientific problems in plasma physics and fusion research. 
In this context, it is shown that a 12D parameter space can be scanned very efficiently, gaining more than an order of magnitude in computational cost over the standard adaptive approach. 
Moreover, it allows for the uncertainty propagation and sensitivity analysis in higher-dimensional plasma microturbulence problems, which would be almost impossible to tackle with standard screening approaches.
\end{abstract}


\end{frontmatter}


\section{Introduction}
It has become well established that a proper quantification of uncertainty is an important step towards predictive numerical simulations of real-world phenomena. 
Whether stemming from measurement errors, incomplete knowledge or inherent variabilities, uncertainty is intrinsic to most real-world problems and it needs to be accounted for \emph{ab initio}. 

In this paper, we assume the considered phenomenon is model-driven, i.e., represented using a mathematical model, such as a differential or an integral equation.
Quantifying and reducing uncertainties in such problems is done within the framework of \emph{uncertainty quantification} (UQ).
Specifically, we focus on \emph{uncertainty propagation}, also known as forward UQ. 
In forward UQ, the input uncertainty is typically modelled via a probability density function stemming from e.g., estimations from measurement data or expert opinion.
By sampling this density, the underlying model is evaluated once for each sample. 
These samples can be, for example, pseudo-random numbers or deterministic collocation points. 
In this paper, we employ the latter. 
After performing all simulations, the ensemble of model evaluations is used to assess statistics such as expectation and standard deviation.
Note that uncertainty propagation is an example of an \emph{outer-loop} scenario,
i.e., scenarios in which a specific quantity is computed using multiple evaluations of the underlying model (see \cite{PWG18}).
To summarize, we focus on forward UQ in which the modelling of the uncertain parameters is based on expert opinion supported by experimental measurements.

The real-world application under consideration here is the simulation of microturbulence in magnetized fusion plasmas.
This problem is of high practical relevance for efforts such as the ITER\footnote{ITER was initially an acronym for International Thermonuclear Experimental Reactor. Nowadays it is mainly referred to the Latin word ``iter'', meaning ``the way''.} experiment, which will aim at creating -- for the very first time -- a self-sustained ("burning") plasma in the laboratory. 
This amounts to a milestone on the way towards a future fusion power plant. A physics obstacle on this route are small-scale fluctuations which cause energy loss rates despite sophisticated plasma confinements via strong and shaped magnetic fields. This microturbulence is driven by the free energy provided by the unavoidably steep plasma temperature and density gradients. 
Unfortunately, the measurement of these gradients, as well as further decisive physics parameters affecting the underlying microinstabilities are subject to uncertainties, thus requiring a UQ framework. 
In this paper, we employ the established plasma microturbulence simulation code {\sc Gene} \cite{Goe11, Je00} and focus on linear gyrokinetic eigenvalue problems in 5D phase space. 
Even without the nonlinear terms, the significant computational requirements and large number of stochastic parameters, typically referred to as the \emph{curse of dimensionality} (see, e.g., \cite{BG04}), render the quantification of uncertainty in plasma problems challenging. 

Overcoming or delaying the curse of dimensionality is one of the grand challenges in UQ and in scientific computing in general. 
Approximations based on sparse grids \cite{BG04} have been well established as suitable counter-measures. 
To this end, starting with works such as \cite{MZ09, NTW08, XH05}, sparse grid approximations have been extensively used in UQ. 
In recent years, significant efforts have been made towards designing enhanced approximation strategies for computationally expensive stochastic problems. 
A non-exhaustive list includes \cite{CM13, CEP12}, where the so-called sparse pseudo-spectral projection (PSP) method was proposed and analysed. 
With PSP, one constructs sparse approximations free of internal aliasing error by choosing the maximum degree of the projection basis such that the continuous and discrete inner products coincide when applied to the basis polynomials. 
Moreover, in \cite{CM13}, several adaptive strategies were proposed. 
In \cite{Wi16}, two adaptive methods -- a nested approach and a PSP approximation with directional refinement -- were studied in problems having multiple design and stochastic parameters. 
In addition, there is a growing interest in employing UQ in the numerical simulation of fusion plasmas. 
In \cite{Goe14, To13}, a simple uniform, deterministic parameter scan was used in both linear and nonlinear gyrokinetic simulations. 
In both papers, the focus was on assessing the sensitivity to changes in the ions temperature gradient to validate the underlying simulation codes. 
Furthermore, in \cite{VH18}, nonintrusive collocation methods were used to quantify uncertainty in a drift-wave turbulence study from the CSDX linear plasma experiment.
However, most existing UQ studies in numerical fusion plasma problems sample the underlying stochastic space using either dense grids, hence suffering from the curse of dimensionality, or a priori chosen sparse grids, whose number of points still grows prohibitively large with the dimensionality.
Moreover, most existing adaptive sparse grid strategies employ deterministic or global information, thus not discriminating between individual directions nor exploiting the (usually) available sensitivity information.
In addition, in most applications the intrinsic stochastic dimensionality is smaller than the given stochastic dimensionality, i.e., only a subset of stochastic inputs is important.

To this end, in this paper we formulate and test a novel structure exploiting adaptive sparse approximation methodology, as follows.
We build our strategy on the dimension-adaptive algorithms of \cite{GG03, He03}, in which the underlying sparse grid is constructed adaptively via a judiciously chosen linear combination of full subspaces with low-cardinality.
However, we drive the adaptive process using sensitivity information to preferentially refine the directions rendered important.
Specifically, we introduce a \emph{sensitivity scoring system} to assess the importance of the individual stochastic input parameters as well as of their interaction.
To obtain these scores, we perform Sobol' decompositions in each subspace and compute variance surpluses which reveal the importance of each input and of their interaction.
We note that the proposed approach is generic, in the sense that it can be formulated in terms of arbitrary approximation operators and point sets. 
In particular, we consider Lagrange interpolation (see, e.g., \cite{BT04}), which requires the underlying model solution to be continuous, and PSP (see, e.g., \cite{Wi16, Xi10}), which requires square-integrability. 
Moreover, to define these two approximations, we employ two (L)-Leja point constructions (see, e.g., \cite{NJ14}).
We remark that besides the formulation of a novel adaptive sparse approximation strategy for stochastic computations, another novelty of this paper is the undertaking, to the best of our knowledge, of one of the first UQ studies in plasma microinstability analysis.
In addition, the proposed methodology is model-agnostic, provided that certain smoothness assumptions such as continuity or square-integrability are fulfilled.

The remainder of this paper is organized as follows. 
In Section \ref{sec:background}, we introduce our notation and provide a brief background on plasma microturbulence simulation and dimension-adaptive sparse grids, on which we construct our proposed approach based on sensitivity scores.
We present in detail our proposed algorithm in Section \ref{sec:methodology}.
We present our numerical results in two plasma mictroturbulence test cases in Section \ref{sec:results}.
In Section \ref{subsec:cbc_test_case}, we consider a modified version of the benchmark \cite{Di00} with eight stochastic inputs. 
In Section \ref{subsec:AUG}, we investigate a more realistic test case based on the study from \cite{Fr18} with three or 12 uncertain inputs. 
Finally, we summarize and give an outlook in Section \ref{sec:conclusions}.

\section{Background} \label{sec:background}

\subsection{Problem formulation} \label{subsec:prob_form}
In this work, we are interested in the quantification of uncertainty of complex, real-world phenomena such as plasma microturbulence analysis.
These problems are typically governed by a model $\mathcal{F}$ specified by a complex and nonlinear ODE/PDE system.
However, the solution to this model is very rarely, if ever, available analytically, hence we resort to numerical approximations.
Let us denote the discretized version of the underlying continuous model by $\mathcal{F}_h$ such that $\mathcal{E}(\mathcal{F} - \mathcal{F}_h)$ can be made arbitrarily small, where $\mathcal{E}$ is a suitable error function. 
We assume $\mathcal{F}_h$ is a bounded and measurable function w.r.t.~the Lebesgue measure and the Borel $\sigma$-algebra on $\mathbb{R}$.
Uncertainty enters $\mathcal{F}_h$ via a vector of $d$ stochastic inputs $\boldsymbol{\theta} := (\theta_1, \theta_2, \ldots, \theta_d)$.
We model $\boldsymbol{\theta}$ as a multivariate random vector with independent components $\theta_i$ defined in a probability space $(\Omega, A, P)$ whose event space is $\Omega$ and is equipped with $\sigma-$algebra $A$ and probability measure $P$.
We assume that $\boldsymbol{\theta}$ is a continuous random vector characterized by a probability density function $\boldsymbol{\pi}$ with a product structure, i.e.,
\begin{equation*}
\boldsymbol{\pi}(\boldsymbol{\theta}) := \prod_{i=1}^d \pi_i(\theta_i)
\end{equation*}
with image $\boldsymbol{X} := \boldsymbol{\pi}(\Omega)$ have a product structure as well
\begin{equation*}
\boldsymbol{X} := \bigotimes_{i=1}^d X_i,
\end{equation*}
where $X_i$ is the image of $\pi_i$.
The independence assumption on the components of $\boldsymbol{\theta}$ can be relaxed if a suitable transformation, e.g., a transport map \cite{Ma16} is employed.
In uncertainty propagation, we are interested in computing quantities of interest such as the expectation $\mathbb{E}[\mathcal{F}_h]$ and standard deviation $\sigma[\mathcal{F}_h] := \sqrt{\mathrm{Var}[\mathcal{F}_h]}$ of the output of $\mathcal{F}_h$.
When a single evaluation of $\mathcal{F}_h$ is computationally expensive, as it is assumed in this work, the computation of $\mathbb{E}[\mathcal{F}_h]$ and $\sigma[\mathcal{F}_h] := \sqrt{\mathrm{Var}[\mathcal{F}_h]}$ become prohibitive.
Addressing the challenges of quantifying uncertainty in computationally expensive, real world problems is the main goal of this work.
\subsection{Plasma microturbulence simulations}\label{sec:plasma}

Fusion devices are subject to plasma microturbulence which is driven by the intrinsically steep density and temperature gradients. 
The associated turbulent transport determines the energy confinement time which in turn is a key parameter for creating a burning plasma. 
Any insight into the nature of plasma microturbulence and ways for avoiding turbulence related confinement degradations are therefore crucial for the design of fusion power plants. 

Unfortunately, even turbulence in fluid systems is difficult to assess and considered one of the most important problems in classical physics. 
In magnetically confined plasmas -- basically very hot but dilute ionized gases -- the problem is further complicated by the low collisionality which renders fluid descriptions insufficient in many situations. 
An appropriate description is therefore given by a kinetic model, i.e., 6D Vlasov equations per species, which are coupled via Maxwell's equations. 
Computing corresponding solutions in complex geometries is challenging even on present-day computational resources.
Even more, performing tasks such as uncertainty propagation which require (large) ensembles of such simulations can quickly become computationally prohibitive.
This therefore calls for using surrogate models.
The following subsections summarize the most popular approach for plasma microturbulence analysis, emphasize the need for UQ in plasma mictroturbulence analysis, and introduce the employed plasma turbulence code.

\subsubsection{The gyrokinetic approach and its applications}

The most popular theory for assessing plasma microturbulence is the so-called gyrokinetic theory \cite{BH07, Kro12}.
Acknowledging a time-scale separation between the fast gyromotion of the particles around the magnetic field lines and typical turbulence time scales, the knowledge of the exact position of the particles along their orbit is irrelevant. 
Gyrokinetics therefore effectively removes the gyrophase information and yields a $5$D system of equations which better fits nowadays computational resources and is considered a valid approach for a wide range of plasma parameters.

Several numerical implementations have been developed over the last two decades and encouraging progress has been made \cite{Gar10, Kro12}.
Early gyrokinetic studies were often limited to restricted physics, e.g., adiabatic electrons and simplified geometries, which could only yield qualitative results and predictions.
However, the complexity has meanwhile dramatically increased and flagship codes aim for quantitative validation with various observables obtained from experimental measurements, see e.g., \cite{WG17} and references therein.
While some codes aim for a full flux-driven setup, where profiles and turbulence are self-consistently developing in response to prescribed heat and particle source, these simulations are usually too costly for regular applications and therefore typically performed with reduced physics. 

An alternative scheme is to use the experimentally determined mean temperature and density profiles as well as the magnetic topology in a given time window as fixed physics inputs to the gyrokinetic codes and compute the resulting turbulent fluctuations.
Naturally, all these physics inputs but also the experimental fluctuation observables can only be measured within some uncertainty. 
This can be quite troublesome since plasma profiles are often found to be quite stiff, i.e., a small increase in inputs such as the gradients may cause large differences in the resulting turbulent heat fluxes. 
To this end, the standard practice is to identify the key parameters affecting the scenario at hand and scan these within the error bars. 
Given the enormous computational efforts associated to high-dimensional parameter scans, the former (identification) step is often being performed without considering the nonlinearities in the Vlasov equation which can take up to $50\%$ of the run time. 
Such linear simulations are still highly relevant to characterize the underlying microinstabilities such as \emph{ion} or \emph{electron temperature gradient} (ITG/ETG) driven modes, \emph{trapped electron} modes (TEMs), \emph{micro-tearing} modes (MTM), and many more. Determining their threshold values or transitions usually provides some guideline for input parameters scans in fully nonlinear simulations. 
However, many of such studies are only considering a subset of stochastic input parameters due to the tremendously large required computational cost.
This clearly provides motivation to develop and apply modern UQ methods in plasma microinstability analysis, which we address in this paper.

\subsubsection{The plasma microturbulence code {\sc Gene}} \label{subsec:gene}
The gyrokinetic solver employed in this publication is the {\sc Gene} code -- one of the first grid-based codes in this field which has now been under continuous development for almost two decades \cite{Je00}. 
{\sc Gene} computes the time evolution of gyrocenter distribution functions on a fixed grid in a 5D (3D-2V) position-velocity space. 
The underlying nonlinear partial differential equations of gyrokinetic theory are solved via a mix of numerical methods also widely used in Computational Fluid Dynamics, including finite difference, spectral, finite element, and finite volume schemes. 
Details are described, e.g., in \cite{Goe11}.

Originally, {\sc Gene} had been restricted to flux-tube simulation domains \cite{Be95}, i.e., thin magnetic-field-line following boxes which allow for highly efficient simulations if the radial correlation lengths of the turbulent fluctuations are small compared to the profile scale lengths. 
In this limit, the radial variations of the profiles (as well as their gradients) can be assumed to be constant across the simulation domain. Consequently, one may safely assume periodic boundary conditions in the directions perpendicular to the magnetic field line and apply spectral methods which greatly simplify operators such as gyro-averages. 
For applications in small devices or with steep profiles, however, locality is not a safe assumption anymore and -- by considering full radial profiles -- periodicity is at least lost in the radial direction. 
Correspondingly modified numerical methods, e.g., finite-differences instead of spectral methods, have been added as another option in {\sc Gene} which since then allows for radially global simulations \cite{Goe11} or full flux-surface simulations \cite{Xa14} if toroidal instead of radial background variations are considered, respectively. 
The field-aligned non-orthogonal coordinate system has, however, been kept to exploit the high anisotropy of plasma turbulence which typically has correlation lengths of several meters along a magnetic field line but only of a few centimeters in the perpendicular plane. 
{\sc Gene} simulations are parallelized by domain decomposition in all five dimensions, typically using MPI \cite{mpi-forum}. 
Fully nonlinear simulations may require at least $10$k CPU-hours which is why linear local (flux-tube) simulations will be employed for testing the UQ framework.
Linear local simulations have a runtime of up to at most a few hours.
These simulations enable the quantification of uncertainty of the microinstability character.
Our long-term goal, however, is to quantify the uncertainty in fully nonlinear, turbulent simulations, which we will address in subsequent publications.

\subsection{Approximation with dimension-adaptive sparse grids} \label{sec:sg_approx}

In this section, we summarize dimension-adaptive sparse grid approximations \cite{GG03, He03}, on which we build our proposed methodology in Section \ref{sec:methodology}.
Let $f : X_i \rightarrow \mathbb{R}$ denote a univariate function and let $\mathcal{U}^{i}[f]$ be a sequence of univariate linear continuous operators depending on the marginals $\pi_i$.
Moreover, let $\mathcal{U}^{i}_{\ell}[f]$ be approximations such that 
\begin{equation*}
\left\| \mathcal{U}^{i}[f] - \mathcal{U}^{i}_{\ell}[f] \right\| \xrightarrow{\ell \rightarrow \infty} 0, \quad i = 1, 2, \ldots, d,
\end{equation*}
in a suitable norm $\left\| \cdot \right\|$, where $1 \leq \ell \in \mathbb{N}$ is referred to as \emph{level}.
$\mathcal{U}^{i}_{\ell}[f]$ is obtained from discrete evaluations of $\mathcal{U}^{i}[f]$ at $m(\ell)$ points in $X_i$, where $m(\ell) : \mathbb{N} \rightarrow \mathbb{N}$ is called the \emph{level-to-nodes} function.
In this work, $\mathcal{U}^{i}[f]$ is interpolation or PSP.
In general, we assume that $\mathcal{U}^{i}[f]$ are \emph{global} operators, i.e., their support is the entire domain $X_i$.

The key idea behind formulating sparse grid approximations is to make use of the fact that $\mathcal{U}^{i}_{\ell}[f]$ converges to $\mathcal{U}^{i}[f]$ as $\ell \rightarrow \infty$ and write $\mathcal{U}^{i}[f]$ as a \emph{telescoping series} of the form
\begin{equation*}
\mathcal{U}^{i}[f] = \mathcal{U}^{i}_1[f] + (\mathcal{U}^{i}_2[f] - \mathcal{U}^{i}_1[f]) + (\mathcal{U}^{i}_3[f] - \mathcal{U}^{i}_2[f]) + \ldots = \sum_{\ell = 1}^{\infty} \mathcal{U}^{i}_{\ell}[f].
\end{equation*}
We denote by $\Delta^i_{\ell}[f] := \mathcal{U}^{i}_{\ell}[f] - \mathcal{U}^{i}_{\ell-1}[f]$ the \emph{hierarchical surpluses}, with the convention $\mathcal{U}^i_0[f] := 0$. 
Therefore, the above telescoping series becomes
\begin{equation} \label{eq:telescope_1D}
\mathcal{U}^{i}[f] = \sum_{\ell = 1}^{\infty} \Delta^{i}_{\ell}[f].
\end{equation}

We are interested in approximations that depend on $d \geq 2$ stochastic parameters. 
A natural way to lift the $1$D operators to $d$ dimensions is via \emph{tensorization}, i.e., 
\begin{equation} \label{eq:tensor}
\boldsymbol{\mathcal{U}}^{\boldsymbol{d}}[\mathcal{F}_h] = \big(\mathcal{U}^{1} \otimes \mathcal{U}^{2} \otimes \ldots \otimes \mathcal{U}^{d}\big)[\mathcal{F}_h] = \big(\bigotimes_{i=1}^{d} \mathcal{U}^{i}\big)[\mathcal{F}_h].
\end{equation}
Plugging the telescoping series \eqref{eq:telescope_1D} into the tensorized representation \eqref{eq:tensor}, we obtain
\begin{equation} \label{eq:tensor_delta}
\begin{split}
\boldsymbol{\mathcal{U}}^{\boldsymbol{d}}[\mathcal{F}_h] = & \big(\sum_{\ell_1 = 1}^{\infty} \Delta^{1}_{\ell_1} \otimes \sum_{\ell_2 = 1}^{\infty} \Delta^{2}_{\ell_2} \otimes \ldots \otimes \sum_{\ell_{d} = 1}^{\infty} \Delta^{{d}}_{\ell_{d}}\big)[\mathcal{F}_h] \\ = & \big(\sum_{\boldsymbol{\ell} \in \mathbb{N}^{d}} \Delta^{1}_{\ell_1} \otimes \Delta^{2}_{\ell_2} \otimes \ldots \otimes \Delta^{{d}}_{\ell_{d}}\big)[\mathcal{F}_h] =  \sum_{\boldsymbol{\ell} \in \mathbb{N}^{d}} \big(\bigotimes_{i=1}^{d} \Delta^{i}_{\ell_i}\big)[\mathcal{F}_h] =  \sum_{\boldsymbol{\ell} \in \mathbb{N}^{d}} \boldsymbol{\Delta}^{\boldsymbol{d}}_{\boldsymbol{\ell}}[\mathcal{F}_h],
\end{split}
\end{equation}
where $\boldsymbol{\ell} = (\ell_1, \ell_2, \ldots \ell_{d}) \in \mathbb{N}^{d}$ denotes a \emph{multiindex} and
\begin{equation*}
\boldsymbol{\Delta}^{\boldsymbol{d}}_{\boldsymbol{\ell}}[\mathcal{F}_h] = \sum_{\boldsymbol{z} \in \{0, 1\}^{d}} (-1)^{|\boldsymbol{z}|_1} \boldsymbol{\mathcal{U}}^{\boldsymbol{d}}_{\boldsymbol{\ell}}[\mathcal{F}_h],
\end{equation*}
where $\boldsymbol{\mathcal{U}}^{\boldsymbol{d}}_{\boldsymbol{\ell}}[\mathcal{F}_h] = \big(\bigotimes_{i=1}^{d} \mathcal{U}^{i}_{\ell_i}\big)[\mathcal{F}_h]$ and $|\boldsymbol{z}|_1 := \sum_{i=1}^{d} z_i$.
However, \eqref{eq:tensor_delta} is a representation of $\boldsymbol{\mathcal{U}}^{\boldsymbol{d}}[\mathcal{F}_h]$ with an infinite number of terms, thus unsuitable for computations.
Therefore, we restrict the multiindices $\boldsymbol{\ell}$ to a finite set $\mathcal{L} \subset \mathbb{N}^{d}$ and define
\begin{equation} \label{eq:tensor_delta_finite}
\boldsymbol{\mathcal{U}}^{\boldsymbol{d}}_{\mathcal{L}}[\mathcal{F}_h] = \sum_{\boldsymbol{\ell} \in \mathcal{L}} \boldsymbol{\Delta}^{\boldsymbol{d}}_{\boldsymbol{\ell}}[\mathcal{F}_h].
\end{equation}
$\mathcal{L}$ must be constructed such that the summation in \eqref{eq:tensor_delta_finite} telescopes correctly. 
Such suitable sets are called \emph{admissible} or \emph{downward closed} (see \cite{GG03}).
In particular, for an admissible set $\mathcal{L}$ it holds that $\boldsymbol{\ell} \in \mathcal{L} \Rightarrow \boldsymbol{\ell} - \boldsymbol{e}_i \in \mathcal{L}$ for $i = 1, 2, \ldots, {d}$, where $\boldsymbol{e}_i$ denotes the $i$th unit vector in $\mathbb{R}^{d}$. 
Note that \eqref{eq:tensor_delta_finite} can be rewritten as a \emph{combination scheme} (see, e.g., \cite{GSZ92, RG18}),
\begin{equation} \label{eq:combi_scheme}
\boldsymbol{\mathcal{U}}^{\boldsymbol{d}}_{\mathcal{L}}[\mathcal{F}_h] = \sum_{\boldsymbol{\ell} \in \mathcal{L}} a_{\boldsymbol{\ell}} \boldsymbol{\mathcal{U}}^{\boldsymbol{d}}_{\boldsymbol{\ell}}[\mathcal{F}_h],
\end{equation}
where $a_{\boldsymbol{\ell}} = \sum_{\boldsymbol{z} \in \{0, 1\}^{d}} (-1)^{|\boldsymbol{z}|_1} \chi_{\mathcal{L}}(\boldsymbol{\ell} + \boldsymbol{z})$ and $\chi_{\mathcal{L}}$ is the characteristic function on $\mathcal{L}$, i.e.,  $\chi_{\mathcal{L}}(\boldsymbol{\ell}) = 1$ if $\boldsymbol{\ell} \in \mathcal{L}$ and $\chi_{\mathcal{L}}(\boldsymbol{\ell}) = 0$ otherwise.
Since \eqref{eq:combi_scheme} is computationally more convenient, we use it throughout our numerical experiments.

\subsubsection{Approximation operators}
\label{subsec:1D_operators}
Without loss of generality, we employ the same approximation operators in all $d$ directions in \eqref{eq:combi_scheme}.
Therefore, for simplicity we drop the superscript $i$ in $\mathcal{U}^{i}$ and $\mathcal{U}^{i}_\ell$ and use the notation $\mathcal{U}$ and $\mathcal{U}_\ell$ instead.
In this work, we consider two approximation operators, interpolation and PSP. 
Let $\mathbb{P}_{P_\ell}$ be the space of univariate polynomials of degree $P_\ell \in \mathbb{N}$. 
Further, let $C^0(X_i)$ denote the space of continuous functions $f: X_i \rightarrow \mathbb{R}$ and $L^2(X_i)$ be the separable Hilbert space of square-integrable functions $f: X_i \rightarrow \mathbb{R}$.
To distinguish between interpolation and PSP, we use the superscripts \emph{in} and \emph{psp}.
In addition, to refer to either one of the two approximations, we use the superscript $\emph{op}$.

A popular interpolation approach used in UQ is Lagrange interpolation (see, e.g., \cite{BT04, CEP12, Fo13, NJ14, NTW08}), which we also employ in this work. 
The univariate Lagrange interpolation operator is defined as:
\begin{equation} \label{eq:interp_1D}
\mathcal{U}_\ell^{\mathrm{in}} : C^0(X_i) \rightarrow \mathbb{P}_{P_\ell}, \quad \mathcal{U}_\ell^{\mathrm{in}}[f] := \sum_{n = 1} ^ {m(\ell)} f(\theta_n)L_n(\theta),
\end{equation}
where $\{\theta_n\}_{n=1}^{m(\ell)}$ are interpolation nodes computed w.r.t.~the density $\pi_i$ and $\{L_n(\theta)\}_{n=1}^{m(\ell)}$ are Lagrange polynomials of degree $n - 1$ satisfying the interpolation condition $L_n(\theta_m) = \delta_{nm}$, where $\delta_{nm}$ is Kronecker's delta function. 
For improved numerical stability, we implement \eqref{eq:interp_1D} in terms of the barycentric formula (see, e.g., \cite{BT04}).

Another commonly used approximation operation in UQ is PSP (see, e.g., \cite{CM13, CEP12, Xi10}). 
The PSP operator is defined as a series expansion of the form: 
\begin{equation}\label{eq:spectral_proj_1D}
\mathcal{U}_\ell^{\mathrm{psp}} :L^2(X_i]) \rightarrow \mathbb{P}_{P_\ell}, \quad \mathcal{U}_\ell^{\mathrm{psp}}[f] := \sum_{p = 0} ^ {P_{\ell}} c_p \phi_p(\theta),
\end{equation}
where $\{\phi_p\}_{p=0}^{P_{\ell}}$ are orthonormal polynomials satisfying
\begin{equation*}
\left\langle \phi_p, \phi_q \right\rangle := \int_{X_i} \phi_p(\theta)\phi_q(\theta) \pi(\theta) \mathrm{d}\theta = \delta_{pq}.
\end{equation*}
Moreover, $c_p$ are the \emph{pseudo-spectral coefficients} defined via projection and computed as:
\begin{equation}\label{eq:spectral_coeff}
c_p := \int_{X_i} f(\theta)\phi_p(\theta) \pi_i(\theta) \mathrm{d}\theta \approx \sum_{n=1}^{m(\ell)} f(\theta_n)\phi_p(\theta_n) w_n,
\end{equation}
where $\{\theta_n\}_{n=1}^{m(\ell)}$ are quadrature nodes and $w_n$ are normalized weights, i.e. $\sum_{n = 1} ^ {m(\ell)} w_n = 1$ computed w.r.t.~$\pi_i$.
In this work, the quadrature rule used to compute the coefficients  $c_p$ in \eqref{eq:spectral_coeff} is exact\footnote{Up to the employed arithmetic precision} for polynomials of degree at most $m(\ell) - 1$. 
With this in mind, we choose $P_{\ell}$ in \eqref{eq:spectral_proj_1D} such that $\int_0^1 \phi_p(\theta)\phi_q(\theta) \mathrm{d}u = \sum_{n = 1}^{m(\ell)} \phi_p(\theta_n)\phi_q(\theta_n) w_n$ for all $p, q\leq P_{\ell}$,  that is, there is no internal aliasing error in the underlying projection space (see \cite{CM13, CEP12, Wi16}).
The maximal degree that guarantees this is $P_{\ell} = \left \lfloor{(m(\ell) - 1)/2}\right \rfloor$.
No internal aliasing error in the projection spaces eliminates potential quadrature errors that drastically affect the approximation quality of PSP, as has been numerically shown in \cite{CEP12} and later proved in \cite{CM13}.
\subsubsection{(L)-Leja sequences} \label{subsec:weighted_leja}
In this section, we summarize weighted (L)-Leja sequences, which we use to construct the dimension-adaptive sparse grid interpolation and PSP.
Since these approximations are constructed using tensorizations of one dimensional operators, it is sufficient to show how Leja points are constructed in $1$D w.r.t.~the marginal density $\pi_i$ for $i = 1, 2, \ldots, d$.

The choice of point sets to construct sparse grid interpolation or PSP for uncertainty propagation has a critical impact on the overall computational cost, since for each grid point an evaluation of the forward model is needed.
When these evaluations are computationally expensive, we seek a point set that: 
\begin{itemize}
\item is nested -- so that we can reuse computations from previous levels;
\item grows slowly with the level -- so that $m(\ell)$ is not very large;
\item has good approximation properties -- so that we obtain accurate approximations.
\end{itemize}
A point set having all aforementioned properties is the weighted (L)-Leja sequence (see, e.g., \cite{JWZ18}), which we consider in this paper. 

For interpolation, we employ standard weighted (L)-Leja points constructed as:
\begin{equation}\label{eq:line_leja_points}
\begin{split}
& \theta_1^{\mathrm{in}} = \underset{\theta \in X_i}{\mathrm{argmax}} \ \pi_i(\theta) \\
\ & \theta_n^{\mathrm{in}} = \underset{\theta \in X_i}{\mathrm{argmax}} \ \pi_i(\theta)\prod_{m=1}^{n-1}|(\theta - \theta_m^{\mathrm{in}})|, \quad n = 2, 3, \ldots
\end{split}
\end{equation}
Note that the above point sequence is in general not uniquely defined, because \eqref{eq:line_leja_points} might have multiple solutions. 
In that case, we simply pick one of the maximizers.
In addition, \eqref{eq:line_leja_points} indicates that to increase the interpolation degree from $j-1$ to $j$, we need to add only $\theta_{j}^{\mathrm{in}}$ to $\{\theta_{1}^{\mathrm{in}}, \theta_{2}^{\mathrm{in}}, \ldots, \theta_{j}^{\mathrm{in}}, \}$.
Therefore, we add one new (L)-Leja point per level, i.e., $m(\ell) = \ell$.

For PSP, we need a quadrature node sequence to evaluate the PSP coefficients (see \eqref{eq:spectral_coeff}).
When the input density is symmetric and compactly supported, which is the case, for example, for uniform densities which we consider in this work, \eqref{eq:line_leja_points} places the first (L)-Leja point in the center of the domain, while the next two points are placed on the boundary of the weight function's support.
In this way, adding only one (L)-Leja point at a time will lead to a zero quadrature weight at level $\ell=2$, causing the employed adaptive algorithms to stop prematurely.
Therefore, for PSP, we employ symmetrized (L)-Leja points \cite{SS13}:
\begin{equation} \label{eq:symm_Leja}
\begin{split}
& \theta_{1}^{\mathrm{psp}} := \underset{\theta \in X_i}{\mathrm{argmax}} \ \pi_i(\theta) \\
\ & \theta_{2n}^{\mathrm{psp}} := \underset{\theta \in X_i}{\mathrm{argmax}} \ |\prod_{\ell=1}^{2n-1}(\theta - \theta_{\ell}^{ \mathrm{psp}})|, \quad \theta_{2n + 1}^{\mathrm{qu}} = 2\theta_{1}^{\mathrm{psp}} - \theta_{2n}^{\mathrm{psp}}, \quad n = 1, 2, \ldots.
\end{split}
\end{equation}
In this construction, we add two additional points per level. 
The first point is obtained via the standard (L)-Leja construction \eqref{eq:line_leja_points} and the second is its symmetric point w.r.t.~$\theta_{1}^{\mathrm{psp}}$. 
Therefore, the level-to-nodes mapping is $m(\ell) = 2\ell - 1$.
\begin{rmk}
The symmetrized Leja construction \eqref{eq:symm_Leja} is designed for densities $\pi_i$ which are symmetric w.r.t.~$\theta_{1}^{\mathrm{psp}}$.
If $\pi_i$ is not symmetric w.r.t.~$\theta_{1}^{\mathrm{psp}}$, we can instead employ, for example, standard (L)-Leja points \eqref{eq:line_leja_points} with level-to-nodes mapping $m(\ell) = 2\ell - 1$.
\end{rmk}
For more details on weighed (L)-Leja points and their properties, we refer to \cite{JWZ18, NJ14}.

\subsubsection{Standard dimension-adaptivity} \label{subsec:std_da}
To define the sparse grid approximation \eqref{eq:combi_scheme}, we need an approximation operator (in this work, this is interpolation or PSP), a (discrete) point set to compute these approximations (for this purpose, we employ weighted (L)-Leja points) and a finite multiindex set, $\mathcal{L}$.
To determine $\mathcal{L}$, we employ dimension-adaptivity \cite{GG03, He03}.
In the following, we summarize the basics of this algorithm and refer the reader to \cite{CM13, GG03, He03, NJ14, Wi16} for more details.

In the dimension-adaptive approach, the multiindex set $\mathcal{L}$ is split into two subsets, $\mathcal{O}$ (the old-index set) and $\mathcal{A}$ (the active set) such that $\mathcal{L} := \mathcal{O} \cup \mathcal{A}$ is admissible. 
$\mathcal{O}$ comprises the already visited multiindices, while the multiindices in $\mathcal{A}$ are used to drive the adaptive process. 
In the first step, $\mathcal{O} = \{\boldsymbol{1}_d\}$, because the corresponding number of points is $1^{d} = 1$, and $\mathcal{A} = \{\boldsymbol{1}_d + \boldsymbol{e}_i, i = 1, 2, \ldots, {d} \}$. 
In the remaining steps, the algorithm employs the following principle: if a multiindex $\boldsymbol{\ell} \in \mathcal{A}$ contributes significantly to the current solution, its adjacent neighbours are likely to contribute as well. 
To this end, based on a refinement indicator $\epsilon(\boldsymbol{\cdot})$, $\epsilon(\boldsymbol{\ell})$ is computed for each $\boldsymbol{\ell} \in \mathcal{A}$. 
Afterwards, the multiindex with the largest refinement indicator is moved to $\mathcal{O}$ and all its forward neighbors that preserve the admissibility of $\mathcal{L}$ are added to the active set $\mathcal{A}$. 
Additionally, a surrogate $\rho := \sum_{\boldsymbol{\ell} \in \mathcal{A}} \epsilon(\boldsymbol{\ell})$ of the global error is computed at each step. 
Note that although $\rho$ is a heuristic, it was proven to be an acceptable surrogate of the global error in a large number of numerical studies (see, e.g., \cite{CM13, GG03, Wi16}). 
The adaptivity stops if $\rho < tol^{\mathrm{op}}$, for a user-defined tolerance $tol^{\mathrm{op}}$, if $\mathcal{A} = \emptyset$, or if a user-defined maximum level, $L_{\mathrm{max}}^{\mathrm{op}}$, is reached. 

The essential ingredient in dimension adaptivity is the refinement indicator $\epsilon^{\mathrm{op}}(\boldsymbol{\cdot})$. 
A standard indicator (see \cite{CM13}) which we also employ in this work reads
\begin{equation}\label{eq:std_error_indicator}
\epsilon^{\mathrm{op}}(\boldsymbol{\ell}) := \| \boldsymbol{\Delta}_{\boldsymbol{\ell}}^{\mathrm{op}}[\mathcal{F}_h] \|_{L^2}/C_{\boldsymbol{\ell}}^{\mathrm{op}},
\end{equation}
where $C_{\boldsymbol{\ell}}^{\mathrm{op}}$ is the cost, usually the number of grid points, needed to compute $\boldsymbol{\Delta}_{\boldsymbol{\ell}}^{\mathrm{op}}[\mathcal{F}_h]$. 
In Section \ref{sec:methodology}, we explain why we use the $L^2$ norm of the surplus $\boldsymbol{\Delta}_{\boldsymbol{\ell}}^{\mathrm{op}}[\mathcal{F}_h]$ in \eqref{eq:std_error_indicator} and how does this lead to our proposed refinement indicator based on sensitivity scores.

\subsection{Computing quantities of interest} \label{sec:postproc}
Our goal in this work is to propagate uncertainty in complex, computationally expensive real world problems such as plasma microturbulence analysis. 
To this end, we employ dimension-adaptive sparse grid approximations, summarized in Section \ref{sec:sg_approx}.
Although these approximations yield a surrogate for the forward model, $\mathcal{F}_h$, in uncertainty propagation we are interested in computed quantities such as the expectation, standard deviation or Sobol' indices for sensitivity analysis corresponding to $\mathcal{F}_h$.
In this section, we summarize how these quantities can be estimated at no additional cost from PSP coefficients. 
In Section \ref{sec:methodology}, we show how PSP coefficients are computed for interpolation as well.

Let 
\begin{equation*}
\boldsymbol{\mathcal{U}}^{\mathrm{psp}}[\mathcal{F}_h] := \sum_{\boldsymbol{p} \in \mathcal{P}^{\mathrm{psp}}} c_{\boldsymbol{p}} \boldsymbol{\Phi_p}(\boldsymbol{\theta})
\end{equation*}
denote a multivariate PSP approximation, where $\mathcal{P}^{\mathrm{psp}}$ is a set containing the multivariate PSP polynomial degrees.
For example, if dimension-adaptive sparse grid PSP based on symmetrized (L)-Leja points is employed (see Section \ref{sec:sg_approx}), then
\begin{equation*}
\mathcal{P}^{\mathrm{psp}} = \{\boldsymbol{p} \in \mathbb{N}^{d} : \boldsymbol{0} \leq \boldsymbol{p} \leq \boldsymbol{p}_{\mathrm{max}}  \},
\end{equation*}
where $\boldsymbol{p}_{\mathrm{max}} = (\ell_{1,\mathrm{max}} - 1, \ell_{2, \mathrm{max}} - 1, \ldots, \ell_{d, \mathrm{max}} - 1)$, since for PSP the univariate degrees are $P_{\ell_i} = \left \lfloor{(m(\ell_i) - 1)/2}\right \rfloor = \left \lfloor{(2\ell_i - 2)/2}\right \rfloor = \ell_i - 1, \quad i = 1, 2, \ldots, d$, and $(\ell_{1, max}, \ell_{2, \mathrm{max}}, \ldots, \ell_{d, \mathrm{max}})$ is the maximum reached multiindex by the dimension-adaptive algorithm.

In \cite{Su08, Xi10} it was shown that the expectation of the forward model, $\mathbb{E}[\mathcal{F}_h]$, its standard deviation, $\sigma[\mathcal{F}_h]$, and total Sobol' indices for variance-based global sensitivity analysis, $S_i^T$, can be estimated from the pseudo-spectral coefficients. 
We have
\begin{align}
\label{eq:compute_qoi_psp_coeff}
\mathbb{E}[\mathcal{F}_h] \approx \mathbb{\hat{E}}[\mathcal{F}_h] = c_{\boldsymbol{0}} \nonumber   \\ 
\sigma[\mathcal{F}_h] \approx \hat{\sigma}[\mathcal{F}_h] = \sqrt{\sum_{\boldsymbol{p} \in \mathcal{P}^{\mathrm{psp}} \setminus \{\boldsymbol{0}\} } c_{\boldsymbol{p}}^2}  \\
S_i^T \approx \hat{S}_i^T = \frac{\sum_{\boldsymbol{p} \in \mathcal{J}_{i}^{\mathrm{psp}}} c_{\boldsymbol{p}}^2}{\sigma^2[\mathcal{F}_h]} \nonumber ,
\end{align}
where $\mathcal{J}_{i}^{\mathrm{psp}} := \{\boldsymbol{p} \in  \mathcal{P}^{\mathrm{psp}}: \boldsymbol{p}_i \neq 0 \}$.
Total Sobol' indices comprise the total contribution of a stochastic input to the resulting variance, i.e., its individual contribution as well as contributions due to interactions with other inputs.
For more details on Sobol' indices, we refer the reader to \cite{Su08, Xi10}.

\section{Sensitivity-driven adaptive refinement} \label{sec:methodology}
In this section, we present in detail the major algorithmic contribution of this work.
We build our method on the adaptive algorithm of \cite{GG03, He03}.
Our novelty is a context-aware refinement policy based on sensitivity information such that the important directions from a stochastic perspective are preferentially refined.
We show in Section \ref{subsec:sobol_decomp} how we obtain directional variances from Sobol' decompositions.
In Section \ref{subsec:sens_da}, we describe in detail our proposed adaptive strategy based on sensitivity scores. 
Finally, in Section \ref{subsec:illustrative_example}, we showcase the proposed adaptive strategy in a simple example.

\subsection{Directional variances from Sobol' decompositions} \label{subsec:sobol_decomp}

The essential ingredient in the dimension adaptive algorithm \cite{GG03, He03} is the refinement indicator $\epsilon^{\mathrm{op}}(\boldsymbol{\cdot})$.
In this work, we propose a refinement indicator based on sensitivity information: starting from the $L^2$ norm of the surpluses $\boldsymbol{\Delta}_{\boldsymbol{\ell}}^{\mathrm{op}}[\mathcal{F}_h]$, we assess the importance of each stochastic parameter as well as of their interaction.
We first summarize how we obtain sensitivity information for PSP, and then we show how to connect interpolation and PSP.

For a multiindex $\boldsymbol{\ell}$ in the active set $\mathcal{A}$, the surplus PSP expansion reads:
\begin{equation}\label{eq:spectral_op_fg_mindex}
\boldsymbol{\Delta}_{\boldsymbol{\ell}}^{\mathrm{psp}}[\mathcal{F}_h] := \sum_{\boldsymbol{p} = \boldsymbol{0}}^{\boldsymbol{P}_{\boldsymbol{\ell}}} \Delta c_{\boldsymbol{p}} \boldsymbol{\Phi}_{\boldsymbol{p}}(\boldsymbol{\theta}),
\end{equation}
where $\boldsymbol{\Phi}_{\boldsymbol{p}}(\boldsymbol{\theta}) := \prod_{i = 1}^d \Phi_{p_i}(\theta_i)$ are the multivariate orthonormal PSP basis polynomials and $\boldsymbol{P}_{\boldsymbol{\ell}} := (\left \lfloor{(2\ell_1 - 2)/2}\right \rfloor, \left \lfloor{(2\ell_2 - 2)/2}\right \rfloor, \ldots, \left \lfloor{(2\ell_d - 2)/2}\right \rfloor) = (\ell_1 - 1, \ell_2 - 1, \ldots, \ell_d - 1)$ is the total multivariate degree, since $m(\ell) = 2\ell - 1$ for PSP (cf. Section \ref{subsec:1D_operators}).
Moreover, 
\begin{equation} \label{eq:spectral_op_delta}
\Delta c_{\boldsymbol{p}} := \sum_{\boldsymbol{z} \in \{0, 1\}^{d}} (-1)^{|\boldsymbol{z}|_1} c_{\boldsymbol{p} - \boldsymbol{z}},
\end{equation}
where $\Delta c_{\boldsymbol{0}} := c_{\boldsymbol{0}}$.
From Parseval's identity, we have that
\begin{equation}\label{eq:spectral_l2_norm}
\Big \Vert \boldsymbol{\Delta}_{\boldsymbol{\ell}}^{\mathrm{psp}}[\mathcal{F}_h] \Big \Vert_{L^2}^2 = \sum_{\boldsymbol{p} = \boldsymbol{0}}^{\boldsymbol{P}_{\boldsymbol{\ell}}} \Delta c_{\boldsymbol{p}}^2.
\end{equation}

In \cite{Su08}, it was shown that a PSP expansion is equivalent to a Sobol' decomposition \cite{So00} of the same degree, i.e., the terms in the two decompositions are equivalent.
A Sobol' expansion is used to represent a $d$-variate function as a summation of the function's expectation and variances due to each individual input, which reveal the relative importance of each input direction, and variances stemming from all possible interactions between inputs, which quantify the relative importance of input interactions.
Using the equivalence between PSP and Sobol' decompositions, we can rewrite \eqref{eq:spectral_l2_norm} as
\begin{equation}\label{eq:spectral_op_sobol}
\Big \lVert \boldsymbol{\Delta}_{\boldsymbol{\ell}}^{\mathrm{psp}}[\mathcal{F}_h] \Big \rVert_{L^2}^2 = \sum_{\boldsymbol{p} = \boldsymbol{0}}^{\boldsymbol{P}_{\boldsymbol{\ell}}} \Delta c_{\boldsymbol{p}}^2 = \Delta \mathrm{Var}^{\mathrm{psp}, 0}_{\boldsymbol{\ell}}[\mathcal{F}_h] + \sum_{i=1}^d \Delta \mathrm{Var}^{\mathrm{psp}, i}_{\boldsymbol{\ell}}[\mathcal{F}_h] + \Delta \mathrm{Var}^{\mathrm{psp}, \mathrm{inter}}_{\boldsymbol{\ell}}[\mathcal{F}_h],
\end{equation}
where $\Delta \mathrm{Var}^{\mathrm{psp}, 0}_{\boldsymbol{\ell}}[\mathcal{F}_h] := \Delta c_{\boldsymbol{0}}^2$ is the expectation surplus, usually small (see, e.g., \cite{Wi16}), 
\begin{equation*}
\Delta \mathrm{Var}^{\mathrm{psp}, i}_{\boldsymbol{\ell}}[\mathcal{F}_h] := \sum_{\boldsymbol{p} \in \mathcal{I}^{\mathrm{psp}}_i} \Delta c_{\boldsymbol{p}}^2,
\end{equation*}
are the surplus contributions to the $d$ individual directional variances for $i = 1, 2, \ldots, d$, where 
$\mathcal{I}^{\mathrm{psp}}_i = \{ \boldsymbol{0} < \boldsymbol{p} \leq \boldsymbol{P}_{\boldsymbol{\ell}}: \boldsymbol{p}_i \neq 0 \land \boldsymbol{p}_j = 0,  \forall j \neq i \}$, and $\Delta \mathrm{Var}^{\mathrm{psp}, \mathrm{inter}}_{\boldsymbol{\ell}}[\mathcal{F}_h] := \sum_{\boldsymbol{p} \in \mathcal{P}^{\mathrm{psp}}_{\mathrm{inter}}} \Delta c_{\boldsymbol{p}}^2$, where 
$\mathcal{P}^{\mathrm{psp}}_{\mathrm{inter}} = \bigcup_{i=1}^d \{ \boldsymbol{0} < \boldsymbol{p} \leq \boldsymbol{P}_{\boldsymbol{\ell}}: \boldsymbol{p} \not\in \mathcal{I}_{i}^{\mathrm{psp}} \}$, refers to the surplus variance due to all possible interactions.
Therefore, using the $L^2$ norm of the surplus in the standard indicator \eqref{eq:std_error_indicator} means using stochastic information to drive the adaptive process, which is desirable in UQ contexts.
For more details on Sobol' decompositions and Sobol' indices, we refer to \cite{Fo13, So00, Su08, Wi16} and the references therein.
We can also write $\boldsymbol{\Delta}_{\boldsymbol{\ell}}^{\mathrm{psp}}[\mathcal{F}_h]$ in \eqref{eq:spectral_op_fg_mindex} in terms of a scalar index $j = 1, 2, \ldots, N_{\boldsymbol{\ell}} := \prod_{i = 1}^d \ell_i$ representing the position of $\boldsymbol{p}$ in $\{\boldsymbol{0}, \ldots, \boldsymbol{P}_{\boldsymbol{\ell}}\}$.

For Lagrange interpolation, there is no direct connection to Sobol' expansions. 
To this end, we perform a transformation from the Lagrange basis to the orthonormal basis of the same (multivariate) degree to obtain the equivalent pseudo-spectral coefficients (see, e.g., \cite{Fo13, Ga05}).
Let $\boldsymbol{\mathcal{U}}_{\boldsymbol{\ell}}^{\mathrm{in}}$ denote the full-grid interpolation operator for multiindex $\boldsymbol{\ell}$. 
The change of basis from Lagrange interpolation and PSP is done as:
\begin{equation} \label{eq:spectral_to_interp}
\boldsymbol{\mathcal{U}}_{\boldsymbol{\ell}}^{\mathrm{in}}[\mathcal{F}_h] = \sum_{\boldsymbol{p} = \boldsymbol{0}}^{\boldsymbol{I}_{\boldsymbol{\ell}}} c_{\boldsymbol{p}} \boldsymbol{\Phi}_{\boldsymbol{p}}(\boldsymbol{\theta}) = \sum_{j = 1}^{M_{\boldsymbol{\ell}}} c_{j} \boldsymbol{\Phi}_{j}(\boldsymbol{\theta}),
\end{equation}
where $\{\boldsymbol{\Phi}_{\boldsymbol{p}}(\boldsymbol{\theta})\}_{\boldsymbol{p} = \boldsymbol{0}}^{\boldsymbol{I}_{\boldsymbol{\ell}}}$ is the equivalent PSP basis, $\{c_{\boldsymbol{p}}(\boldsymbol{\theta})\}_{\boldsymbol{p} = \boldsymbol{0}}^{\boldsymbol{I}_{\boldsymbol{\ell}}}$ are the PSP coefficients and $\boldsymbol{I}_{\boldsymbol{\ell}} = (\ell_1 - 1, \ell_2 - 1, \ldots, \ell_d - 1)$, since $m(\ell) = \ell$ for interpolation. 
We also use the scalar index $j = 1, 2, \ldots, {M_{\boldsymbol{\ell}}} := \prod_{i=1}^d \ell_i$. 
To obtain the PSP coefficients $\{c_{j}\}_{j=1}^{M_{\boldsymbol{\ell}}}$, we solve $ \sum_{j = 1}^{M_{\boldsymbol{\ell}}} c_{j} \boldsymbol{\Phi}_{j}(\boldsymbol{\theta}_k) = \boldsymbol{\mathcal{U}}_{\boldsymbol{\ell}}^{\mathrm{in}}[\mathcal{F}_h(\boldsymbol{\theta}_k)]$ for all line (L)-Leja points $\boldsymbol{\theta}_k$ associated to the multiindex $\boldsymbol{\ell}$ (see, e.g., \cite{Fo13}).
Then, we compute the surpluses \eqref{eq:spectral_op_delta} and the squared $L^2$ norm \eqref{eq:spectral_op_sobol}. 
\begin{rmk} \label{re:remark_spectral_interp_basis}
For the employed level-to-nodes mappings, the bases for interpolation and PSP have the same size $M_{\boldsymbol{\ell}} = N_{\boldsymbol{\ell}}$, therefore, the number of pseudo-spectral coefficients is the same in both cases. 
However, for PSP, $m(\ell) = 2\ell - 1$, whereas $m(\ell) = \ell$ for interpolation.
Therefore, to achieve the same accuracy, we expect interpolation to require fewer grid points.
\end{rmk}
\subsection{Sensitivity-driven dimension adaptivity} \label{subsec:sens_da}
Next, we present in detail our proposed adaptive strategy.
In Section \ref{subsubsec:calc_sens_scores}, we describe how to assess what we call \emph{sensitivity scores}.
We depart from standard adaptive techniques which use global information to guide the refinement process and use instead sensitivity information in each subspace to preferentially refine the directions rendered important.
Since several sensitivity scores can be equal, we introduce a classification algorithm in Section \ref{subsubsec:max_sens_scores} to ensure that one maximum score is found per refinement step.
We present the proposed dimension-adaptive algorithm based on sensitivity scores in Section \ref{subsubsec:sens_scores_algo}.
\subsubsection{Sensitivity scores} \label{subsubsec:calc_sens_scores}

The standard error indicator \eqref{eq:std_error_indicator} employed in this work depends on the norm of the surpluses, $\| \boldsymbol{\Delta}_{\boldsymbol{\ell}}^{\mathrm{op}}[\mathcal{F}_h]\|_{L^2}$, which represents global information.
In this way the adaptive algorithm does not discriminate between the individual directions nor their interactions.
Having a finer control over the individual input directions as well as their interaction is especially desired in problems in which these directions are anisotropically coupled.
In addition, in the majority of practical applications, such as plasma mictroturbulence simulation, the \emph{intrinsic stochastic dimensionality} is usually smaller than the total number of stochastic parameters, $d$, i.e., only a subset of inputs are \emph{important} in the uncertainty propagation problem.

To this end, we propose an adaptive approach to better exploit the structure of the underlying problem.
We introduce a \emph{sensitivity scoring system} to quantify the contribution of each subspace to the individual variances as well as to the variance due to the parameters' interaction.
Specifically, for each multiindex $\boldsymbol{\ell}$ in the active set $\mathcal{A}$, we first compute $\| \boldsymbol{\Delta}_{\boldsymbol{\ell}}^{\mathrm{op}}[\mathcal{F}_h]\|_{L^2}^2$ using \eqref{eq:spectral_op_sobol}\footnote{For interpolation, we do the basis transformation \eqref{eq:spectral_to_interp} and compute the squared $L^2$ norm using \eqref{eq:spectral_op_sobol}.}.
Recall that the summation terms in $\| \boldsymbol{\Delta}_{\boldsymbol{\ell}}^{\mathrm{op}}[\mathcal{F}_h]\|_{L^2}^2$ comprise the directional variance surpluses for each input parameter as well as the contribution due to the inputs' interaction. 
Hence, using $d + 1$ user-defined tolerances $\boldsymbol{\tau}^{\mathrm{op}} = (\tau_{1}^{\mathrm{op}}, \tau_{2}^{\mathrm{op}}, \ldots, \tau_{d}^{\mathrm{op}}, \tau_{d + 1}^{\mathrm{op}})$, we compute the \emph{sensitivity score} of the multiindex $\boldsymbol{\ell}$, $s_{\boldsymbol{\ell}}^{\mathrm{op}} \in \mathbb{N}$. 
Initially, $s_{\boldsymbol{\ell}}^{\mathrm{op}} = 0$. 
For $i = 1, 2, \ldots, d$, we increase $s_{\boldsymbol{\ell}}^{\mathrm{op}}$ by one if the individual variance surpluses satisfy $\Delta \mathrm{Var}_{\boldsymbol{\ell}}^{\mathrm{op}, i}[\mathcal{F}_h] \geq \tau_{i}^{ \mathrm{op}}$. 
Hence, after this step, $s_{\boldsymbol{\ell}}^{\mathrm{op}}$ can be at most $d$. 
Finally, if the variance surplus due to the interaction of the stochastic parameters, $\Delta \mathrm{Var}^{\mathrm{op}, \mathrm{inter}}_{\boldsymbol{\ell}}[\mathcal{F}_h]$, satisfies $\Delta V^{\mathrm{op}, \mathrm{inter}}_{\boldsymbol{\ell}}[\mathcal{F}_h] \geq \tau_{d + 1}^{\mathrm{op}}$, we increase $s_{\boldsymbol{\ell}}^{\mathrm{op}}$ by one as well. 
Therefore, $s_{\boldsymbol{\ell}}^{\mathrm{op}}$ can take integer values between $0$ and $d + 1$.

Since the introduced sensitivity scoring system is based on the $d$ individual stochastic input directions as well as their interaction, it filters the important directions and ensures that they are preferentially refined.
Therefore, when the underlying inputs are anisotropically coupled or when the intrinsic stochastic dimensionality is smaller than $d$, the value of the sensitivity score will reflect these properties.
Note, in addition, that our scoring system distinguishes between individual and interaction variance surpluses.
In this way, we ensure that if the mixed directions are stochastically unimportant, we prevent the algorithm to refine too much these directions, which are computationally expensive due to their larger number of grid points.
In summary, when the underlying stochastic problem has a rich structure, the proposed sensitivity scoring system will \emph{explore} and \emph{exploit} that structure.

We summarize the computation of the sensitivity scores in Algorithm \ref{algo:scores_comp}. 
the input probability density, $\boldsymbol{\pi}$, w.r.t.~which all computations are performed, the tolerances, $\boldsymbol{\tau}^{\mathrm{op}}$ and the hierarchical surplus $\boldsymbol{\Delta}_{\boldsymbol{\ell}}^{\mathrm{op}}[\mathcal{F}_h]$.
Note that the choice of $\boldsymbol{\tau}^{\mathrm{op}}$ is problem dependent.
For example, if, based on expert opinion or pre-existing knowledge, certain input parameters or their interaction are known to be more important, the corresponding tolerances in $\boldsymbol{\tau}^{\mathrm{op}}$ should be chosen accordingly.
However, when no such knowledge is available, we recommend a conservative choice in which all components of $\boldsymbol{\tau}^{\mathrm{op}}$ are equal.
\begin{algorithm}
\caption{Sensitivity Scores Computation}\label{algo:scores_comp}
\begin{algorithmic}[1]
\Procedure{ComputeSensitivityScore}{$\boldsymbol{\pi}, \boldsymbol{\tau}^{\mathrm{op}}, \boldsymbol{\Delta}^{\mathrm{op}}_{\boldsymbol{\ell}}[\mathcal{F}_h]$}
\State $s^{\mathrm{op}}_{\boldsymbol{\ell}} := 0$
\State Compute $\| \boldsymbol{\Delta}^{\mathrm{op}}_{\boldsymbol{\ell}}[\mathcal{F}_h] \|_{L^2}^2$ to obtain the variance $\Delta \mathrm{Var}^{\mathrm{op}}_{\boldsymbol{\ell}}[\mathcal{F}_h]$ w.r.t.~$\boldsymbol{\pi}$
\State Decompose $\Delta \mathrm{Var}^{\mathrm{op}}_{\boldsymbol{\ell}}[\mathcal{F}_h]$ via \eqref{eq:spectral_op_sobol} to obtain all unnormalized Sobol' indices
\begin{equation*}
\Delta \mathrm{Var}^{\mathrm{op}}_{\boldsymbol{\ell}}[\mathcal{F}_h] = \sum_{i=1}^{d} \Delta \mathrm{Var}^{\mathrm{op}, i}_{\boldsymbol{\ell}}[\mathcal{F}_h] + \Delta \mathrm{Var}^{\mathrm{op}, \mathrm{inter}}_{\boldsymbol{\ell}}[\mathcal{F}_h]
\end{equation*}
\For{$i\gets 1, 2, \ldots, d$} \label{step:algo1_begin_compute_score}
\If{$\Delta \mathrm{Var}^{\mathrm{op}, i}_{\boldsymbol{\ell}}[\mathcal{F}_h] \geq \tau^{\mathrm{op}}_{i}$}
\State $s^{\mathrm{op}}_{\boldsymbol{\ell}} = s^{\mathrm{op}}_{\boldsymbol{\ell}} + 1$
\EndIf
\EndFor
\If{$\Delta \mathrm{Var}^{\mathrm{op}, \mathrm{inter}}_{\boldsymbol{\ell}}[\mathcal{F}_h] \geq \tau^{\mathrm{op}}_{d + 1}$}
\State $s^{\mathrm{op}}_{\boldsymbol{\ell}} = s^{\mathrm{op}}_{\boldsymbol{\ell}} + 1$ \label{step:algo1_end_compute_score}
\EndIf
\State \Return $s^{\mathrm{op}}_{\boldsymbol{\ell}}$
\EndProcedure
\end{algorithmic}
\end{algorithm}

\subsubsection{Maximum sensitivity score} \label{subsubsec:max_sens_scores}

The adaptive process in our proposed approach is driven by the multiindex from the active set $\mathcal{A}$ with the largest sensitivity score.
To this end, if several subspaces have equal scores we need an additional step to distinguish between them.
Note that two or more subspaces have the same sensitivity score if the same number of directional variances are larger than the prescribed tolerances. 
However, these directions do not need to be necessarily the same. 
With this in mind, we distinguish between subspaces having equal sensitivity scores 
via the following \emph{classification} algorithm.

First, we compute
\begin{equation*}
\Delta \mathrm{Var}_{\boldsymbol{\ell}}^{\mathrm{op}, \mathrm{tot}}[\mathcal{F}_h] := \sum_{i=1}^d \Delta \mathrm{Var}_{\boldsymbol{\ell}}^{\mathrm{op}, i}[\mathcal{F}_h] + \Delta \mathrm{Var}_{\boldsymbol{\ell}}^{\mathrm{op}, \mathrm{inter}}[\mathcal{F}_h],
\end{equation*}
and then select the subspace with the largest $\Delta \mathrm{Var}_{\boldsymbol{\ell}}^{\mathrm{op}, \mathrm{tot}}[\mathcal{F}_h]$.
In this way, if two or more subspaces contribute significantly in the same number of directions, we select the subspace with the largest global contribution.
We summarize the aforementioned strategy for finding the subspace with the maximum score in Algorithm \ref{algo:max_score}. 
The inputs are two sets: $\mathcal{S}$, comprising all current scores, and $\mathcal{D}$, which contains all current surpluses.
In lines \ref{step:algo2_begin_max} -- \ref{step:algo2_end_max}, we find the maximum sensitivity scores.
If only one maximum score exists, then the algorithm ends (lines \ref{step:algo2_one_max_begin} -- \ref{step:algo2_one_max_end}).
In the case when several maximum scores exist, we select the subspace with the largest $\Delta \mathrm{Var}_{\boldsymbol{\ell}}^{\mathrm{op}, \mathrm{tot}}$ (lines \ref{step:algo2_more_max_begin} -- \ref{step:algo2_more_max_end}).
In the unlikely case that $\Delta \mathrm{Var}_{\boldsymbol{\ell}}^ {\mathrm{op}, \mathrm{tot}}$ is the same for several scores, the algorithm returns the multiindex corresponding to the last score.
\begin{algorithm}
\caption{Maximum Sensitivity Score Computation}\label{algo:max_score}
\begin{algorithmic}[1]
\Procedure{FindIndexMaximumScore}{$\mathcal{S}, \mathcal{D}$}
\State $I := []$
\State $\Delta \mathrm{Var}^{\mathrm{op}}_{\mathrm{max}}[\mathcal{F}_h] := 0, \quad \boldsymbol{\ell}^{\mathrm{op}}_{\mathrm{max}} := \boldsymbol{1}$
\State Find the maximum sensitivity scores $s^{\mathrm{op}}_{\boldsymbol{\ell}_n}$ from $\mathcal{S}$ \label{step:algo2_begin_max}
\State Append to $I$ the scalars $n = 1, 2, \ldots, n_{\mathrm{max}}$, where $n_{\mathrm{max}}$ is the number of max scores \label{step:algo2_end_max}
\State $n_{\mathrm{max}} := |I|$
\If {$n_{\mathrm{max}} = 1$} \label{step:algo2_one_max_begin}
\State $p := I[1]$
\State $\boldsymbol{\ell}^{\mathrm{op}}_{\mathrm{max}} := \boldsymbol{\ell}_p$ \label{step:algo2_one_max_end}
\Else
\For{$m\gets 1, 2, \ldots, n_{\mathrm{max}}$} \label{step:algo2_more_max_begin}
\State $q := I[m]$
\State Take $\boldsymbol{\Delta}^{\mathrm{op}}_{\boldsymbol{\ell}_q}[\mathcal{F}_h]$ from $\mathcal{D}$ and compute $\Delta \mathrm{Var}^{\mathrm{op}}_{\boldsymbol{\ell}_q}[\mathcal{F}_h]$
\If{$\Delta \mathrm{Var}^{\mathrm{op}}_{\boldsymbol{\ell}_q}[\mathcal{F}_h] \geq \Delta \mathrm{Var}^{\mathrm{op}}_{\mathrm{max}}[\mathcal{F}_h]$}
\State $\Delta \mathrm{Var}^{\mathrm{op}}_{\mathrm{max}}[\mathcal{F}_h] := \Delta \mathrm{Var}^{\mathrm{op}}_{\boldsymbol{\ell}_q}[\mathcal{F}_h]$
\State $\boldsymbol{\ell}^{\mathrm{op}}_{\mathrm{max}} := \boldsymbol{\ell}_q$ \label{step:algo2_more_max_end}
\EndIf
\EndFor
\EndIf
\State \Return $\boldsymbol{\ell}^{\mathrm{op}}_{\mathrm{max}}$
\EndProcedure
\end{algorithmic}
\end{algorithm}

\subsubsection{Sensitivity-driven dimension-adaptive sparse grid approximations} \label{subsubsec:sens_scores_algo}

\begin{algorithm}
\caption{Sensitivity-driven Dimension-adaptive Sparse Grid Algorithm}\label{algo:main_algo}
\begin{algorithmic}[1]
\Procedure{SensitivityDrivenAdaptiveSparseGridApprox}{$\mathcal{F}_h$, $\boldsymbol{\pi}$, $\boldsymbol{\tau}^{\mathrm{op}}, L^{\mathrm{op}}_{\mathrm{max}}$}
\State $\boldsymbol{1}_d := (1, 1, \ldots, 1)$ \label{step:main_algo_step_1}
\State $\mathcal{O} := \emptyset, \quad \mathcal{A} := \{\boldsymbol{1}_d\}$ \label{step:main_algo_step_2}
\State $\mathcal{S} := \emptyset, \quad \mathcal{D} := \emptyset$ \label{step:main_algo_step_3}
\State $s^{\mathrm{op}}_{\boldsymbol{1}} :=  \textsc{ComputeSensitivityScore}(\boldsymbol{\pi}, \boldsymbol{\tau}^{\mathrm{op}}, \boldsymbol{\Delta}^{\mathrm{op}}_{\boldsymbol{1}}[\mathcal{F}_h])$
\State $\mathcal{S} = \mathcal{S} \cup \{s^{\mathrm{op}}_{\boldsymbol{1}}\}, \quad \mathcal{D} = \mathcal{D} \cup \{\boldsymbol{\Delta}^{\mathrm{op}}_{\boldsymbol{1}}[\mathcal{F}_h]\}$
\While{$\mathrm{all}(\mathcal{S}) \neq 0$ or $\mathcal{A} \neq \emptyset$ or $\max(\mathcal{L}) < L^{\mathrm{op}}_{\mathrm{max}}$}
\State $\boldsymbol{k} := \textsc{FindIndexMaximumScore}(\mathcal{S}, \mathcal{D})$ \label{step:main_algo_step_11}
\State $\mathcal{A} = \mathcal{A} \setminus \{\boldsymbol{k}\}, \quad \mathcal{O} = \mathcal{O} \cup \{\boldsymbol{k}\}$
\State $\mathcal{S} = \mathcal{S} \setminus \{s^{\mathrm{op}}_{\boldsymbol{k}}\}, \quad \mathcal{D} = \mathcal{D} \setminus \{\boldsymbol{\Delta}^{\mathrm{op}}_{\boldsymbol{k}}[\mathcal{F}_h]\}$
\For{$i\gets 1, 2, \ldots, d$}
\State $\boldsymbol{r} \gets \boldsymbol{k} + \boldsymbol{e}_i$
\If{$\boldsymbol{r} - \boldsymbol{e}_q \in \mathcal{O}$ for all $q = 1, 2,  \ldots, d$}
\State $\mathcal{A} = \mathcal{A} \cup \{\boldsymbol{r}\}$
\State $s^{\mathrm{op}}_{\boldsymbol{r}} :=  \textsc{ComputeSensitivityScore}(\boldsymbol{\pi}, \boldsymbol{\tau}^{\mathrm{op}}, \boldsymbol{\Delta}^{\mathrm{op}}_{\boldsymbol{r}}[\mathcal{F}_h])$
\State $\mathcal{S} = \mathcal{S} \cup \{s^{\mathrm{op}}_{\boldsymbol{r}}\}, \quad \mathcal{D} = \mathcal{D} \cup \{\boldsymbol{\Delta}^{\mathrm{op}}_{\boldsymbol{r}}[\mathcal{F}_h]\}$
\EndIf
\EndFor
\EndWhile 
\State $\mathcal{L} = \mathcal{O} \cup \mathcal{A}$
\State Determine the PSP coefficients $\{c_{\boldsymbol{\ell}}\}_{\boldsymbol{\ell} \in \mathcal{L}}$
\State Compute $\mathbb{\hat{E}}[\mathcal{F}_h]$, $\hat{\sigma}[\mathcal{F}_h]$, $\hat{S}_1^T, \hat{S}_2^T, \ldots, \hat{S}_d^T$ from $\{c_{\boldsymbol{\ell}}\}_{\boldsymbol{\ell} \in \mathcal{L}}$ using \eqref{eq:compute_qoi_psp_coeff}
 \label{step:main_algo_last_step}
\State \Return $\mathbb{\hat{E}}[\mathcal{F}_h]$, $\hat{\sigma}[\mathcal{F}_h]$, $\hat{S}_1^T, \hat{S}_2^T, \ldots, \hat{S}_d^T$
\EndProcedure
\end{algorithmic}
\end{algorithm}
We hereby summarize our proposed dimension-adaptive strategy based on sensitivity scores in Algorithm \ref{algo:main_algo}. 
The inputs are the forward model, $\mathcal{F}_h$, the input probability density, $\boldsymbol{\pi}$, the vector $\boldsymbol{\tau}^{\mathrm{op}}$  of $d + 1$ tolerances and the maximum grid level that can be reached, $L_{\mathrm{max}}^{\mathrm{op}}$. 
At lines \ref{step:main_algo_step_1} -- \ref{step:main_algo_step_2}, we initialize $\mathcal{O}$ and $\mathcal{A}$ as in the standard algorithm.
In addition, we initialize two new data structures, $\mathcal{S}$ and $\mathcal{D}$, to store the scores and surpluses for all indices in the active set $\mathcal{A}$ (line \ref{step:main_algo_step_3}).
We proceed by computing the sensitivity score of the first multiiindex in $\mathcal{A}$ using Algorithm \ref{algo:scores_comp} and update $\mathcal{S}$ and $\mathcal{D}$ accordingly. 
Afterwards, in each refinement step, we determine the multiindex with the maximum score (line \ref{step:main_algo_step_11}) based on Algorithm \ref{algo:max_score} and update $\mathcal{A}$ and $\mathcal{O}$. 
Moreover, we also append the largest score and its associated surplus to $\mathcal{S}$ and $\mathcal{D}$, respectively. 
The algorithm continues with adding the forward neighbours of the multindex with the largest sensitivity score provided that the total multiindex set remains admissible.
Further, we assess the sensitivity score for each of these neighbours via Algorithm \ref{algo:scores_comp} and update the sets $\mathcal{O}$, $\mathcal{A}$, $\mathcal{S}$ and $\mathcal{D}$ accordingly. 
Our algorithm terminates if all scores in $\mathcal{S}$ are zero, i.e., if the contributions of all multiindices in $\mathcal{A}$ are rendered stochastically unimportant.
Moreover, the algorithm stops as well if $\mathcal{A} = \emptyset$ or if the maximum level $L_{\mathrm{max}}^{\mathrm{op}}$ is reached.
Note that we do not employ a surrogate for the global error as in the standard dimension-adaptive algorithm of Section \ref{subsec:std_da}, since, first of all, our algorithm relies on error indicators in each subspace. 
Second of all, the algorithm stops, by design, when all subspaces from $\mathcal{A}$ have a zero sensitivity score, i.e., when their variance contribution in all directions becomes insignificant.
At the end, we determine the PSP coefficients $c_{\boldsymbol{\ell}}$ for all $\boldsymbol{\ell} \in \mathcal{L}$ and we use these coefficients to assess the expectation, standard deviation, and total Sobol' indices for global sensitivity analysis via \eqref{eq:compute_qoi_psp_coeff}.

\subsection{Illustrative examples} \label{subsec:illustrative_example}

Before presenting our results in two plasma microturbulence test cases, we first use the proposed methodology in two problems in which the forward model, $\mathcal{F}$, is available analytically.
We use dimension-adaptive sparse grid interpolation and compare our proposed refinement indicator based on sensitivity scores with the standard indicator \eqref{eq:std_error_indicator}.

For a fair comparison, we employ small tolerances ($tol^{\mathrm{in}} = 10^{-8}$ in the standard version and $\boldsymbol{\tau}^{\mathrm{in}} = 10^{-16} \cdot \boldsymbol{1}_{d + 1}$ in our approach) and compute the \emph{$L^2$ approximation error}
\begin{equation} \label{eq:l2_error}
\mathcal{E}^2(\mathcal{F} - \boldsymbol{\mathcal{U}}_{\mathcal{L}}[\mathcal{F}]) := \sqrt{\int_{\boldsymbol{X}} \big(\mathcal{F}(\boldsymbol{\theta}) - \boldsymbol{\mathcal{U}}_{\mathcal{L}}[\mathcal{F}(\boldsymbol{\theta})]\big)^2 \mathrm{d}\boldsymbol{\theta}}
\end{equation}
and the \emph{relative error} of the expectation approximation
\begin{equation} \label{eq:relative_error}
\mathcal{E}_{\mathrm{rel}}(\mathbb{E}[\mathcal{F}] - \hat{\mathbb{E}}[\mathcal{F}]) := |1 - \mathbb{E}[\mathcal{F}]/\hat{\mathbb{E}}[\mathcal{F}]|,
\end{equation}
where $\hat{\mathbb{E}}[\mathcal{F}]$ is estimated using the first PSP coefficient as shown in \eqref{eq:compute_qoi_psp_coeff}.

First, we consider a five-dimensional model $\mathcal{F} : [0, 1]^5 \rightarrow \mathbb{R}$, 
\begin{equation} \label{eq:illustrative_case_1}
\mathcal{F}(\boldsymbol{\theta}) =  1 + \cos{(\pi + 1.5 \theta_1 + 0.5 \theta_2 + 0.05 \theta_3 + 0.1 \theta_4 + 0.002 \theta_5)},
\end{equation}
in which $\boldsymbol{\theta}$ is uniformly distributed in the $5$D hypercube, i.e., $\boldsymbol{\pi}(\boldsymbol{\theta}) = U(0, 1)^5$.
We estimate the five total Sobol' indices as in \eqref{eq:compute_qoi_psp_coeff} using $10^5$ Gauss-Legendre nodes and obtain
\begin{equation*}
\hat{S}_1^T = 0.9034, \quad \hat{S}_2^T = 0.0098, \quad \hat{S}_3^T = 9.8216 \cdot 10^{-4}, \quad \hat{S}_4^T = 3.9287 \cdot 10^{-3}, \quad \hat{S}_5^T = 1.5714 \cdot 10^{-6}.
\end{equation*}
Therefore, the first stochastic parameter is significantly more important that all other four, while $\theta_5$ is the least important parameter.
Since the $L^2$ error $\mathcal{E}^2(\mathbb{E}[\mathcal{F}] - \hat{\mathbb{E}}[\mathcal{F}])$ \eqref{eq:l2_error} and relative error $\mathcal{E}_{\mathrm{rel}}(\mathbb{E}[\mathcal{F}] - \hat{\mathbb{E}}[\mathcal{F}])$ cannot be estimated analytically, we estimate them numerically using $1000$ Monte Carlo samples for the $L^2$ error and the expectation estimate on the Gauss-Legendre grid with $10^5$ points for the relative error.

\begin{figure}[htbp]
  \centering
  \includegraphics[width=0.8\textwidth]{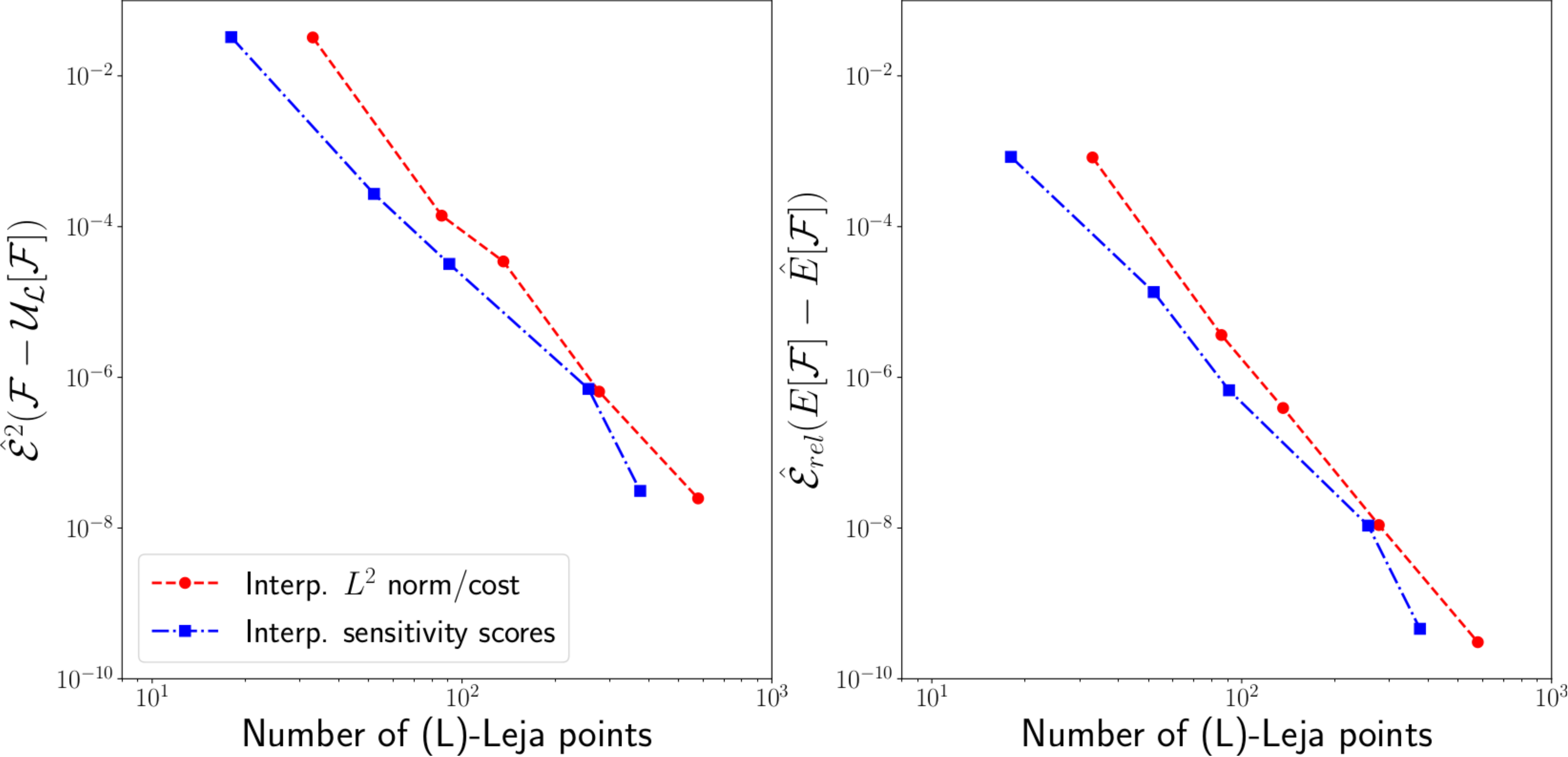}
  \caption{Monte Carlo estimate of the $L^2$ approximation error \eqref{eq:l2_error} (left) and the estimate of the relative error of the expectation approximation \eqref{eq:relative_error} (right) for the example in \eqref{eq:illustrative_case_1}.
  \label{fig:illustrative_ex_5D}}
\end{figure}

We depict the results in Figure \ref{fig:illustrative_ex_5D}.
In the left plot, we visualize the $L^2$ approximation error for the two adaptive schemes and we observe that for similar $L^2$ errors, our approach is cheaper than the standard scheme.
For example, for an $L^2$ error of around $3 \cdot 10^{-8}$, our approach requires $376$ Leja points, whereas the standard approach needs $578$ points.
In the right figure, we depict the estimate for $\mathcal{E}_{\mathrm{rel}}(\mathbb{E}[\mathcal{F}] - \hat{\mathbb{E}}[\mathcal{F}])$ using the same number of Leja points as for the $L^2$ error. 
We observe again that our approach is, for a lower computational cost, as accurate as the standard adaptive method.
Therefore, for the considered $5$D test case \eqref{eq:illustrative_case_1}, we showed that our approach is computationally cheaper than the dimension-adaptive approach based on the standard refinement indicator \eqref{eq:std_error_indicator}.

In general, we expect the proposed refinement indicator based on sensitivity scores to be more accurate than the standard approach in high(er)-dimensional uncertainty propagation problems with anistropically coupled inputs and lower intrinsic dimensionality.
In contrast, in problems with low stochastic dimensionality or problems in which the uncertain inputs are isotropically coupled, we generally do not obtain a significant benefit from using our approach.
We illustrate this point by considering a test case in which we have two uncertain inputs such that (i) one of them is significantly more important that the other and (ii) their interaction is stochastically insignificant.
To this end, consider $\mathcal{F} : [0, 1]^2 \rightarrow \mathbb{R}$, 
\begin{equation} \label{eq:illustrative_case_2}
\mathcal{F}(\boldsymbol{\theta}) =  \cos{(\theta_1 + 0.1 \theta_2)},
\end{equation}
with $\boldsymbol{\pi}(\boldsymbol{\theta}) = U(0, 1)^2$.
The associated total Sobol' indices estimated on a Gauss-Legendre grid comprising $16^2 = 256$ nodes are
\begin{equation*}
\hat{S}_1^T = 0.9908, \quad \hat{S}_2^T = 0.0112,
\end{equation*}
whose values tell us that $\theta_1$ is significantly more important that $\theta_2$.

\begin{figure}[htbp]
  \centering
  \includegraphics[width=0.8\textwidth]{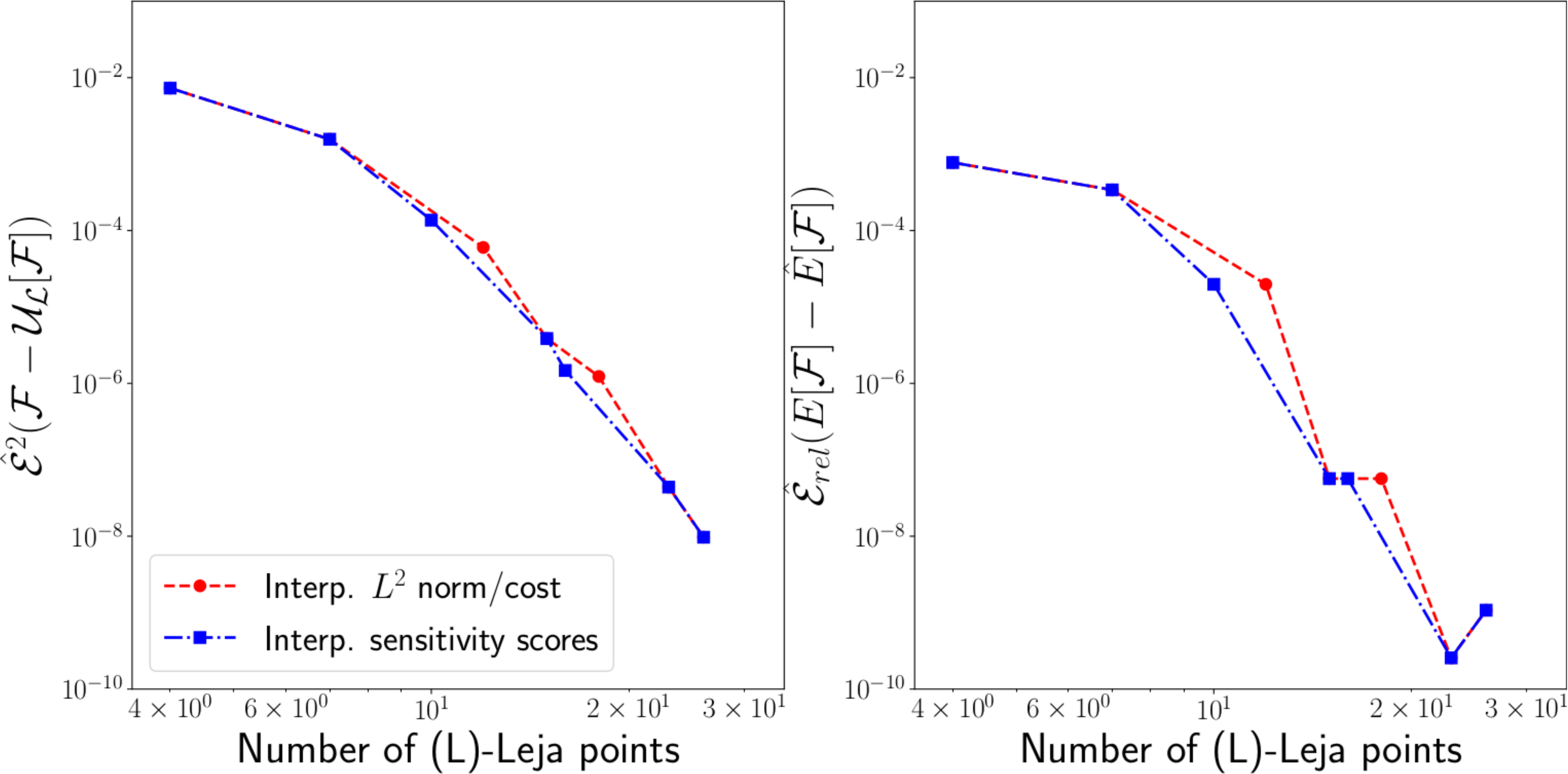}
  \caption{Monte Carlo estimate of the $L^2$ approximation error \eqref{eq:l2_error} (left) and the estimate of the relative error of the expectation approximation \eqref{eq:relative_error} (right) for the example in \eqref{eq:illustrative_case_2}.
  \label{fig:illustrative_ex_2D}}
\end{figure}

The two errors are estimated as in the previous example and depict the results in Figure \ref{fig:illustrative_ex_2D}.
We observe that in this example, our approach yields results of similar accuracy as the standard approach, for a similar cost both from an approximation or expectation estimation perspectives.
Hence, this example, although relatively simplistic, underlines that our proposed sensitivity scores-based approach does not always outperform the standard method.
The behaviour of our scheme depends on the structure of the underlying problem.
Nevertheless, these two examples underline the benefits of dimension-adaptive algorithms in uncertainty propagation: for a problem with five uncertain inputs, we needed at most $576$ Leja points to obtain accurate approximations, whereas for the $2$D examples, at most $32$ points were sufficient to obtain accurate results as well.

\section{Application to plasma microinstability analysis} \label{sec:results}

In this section, we present results of the proposed UQ method applied to two plasma microinstability test cases.
In addition, we also employ the standard adaptive strategy summarized in Section \ref{subsec:std_da}, which we compare with our method.
Both test cases represent linear local (flux-tube) simulations in which microinstabilities are characterised using the linear eigenvalue solver from the gyrokinetic code {\sc Gene} (recall Section \ref{subsec:gene}), hence, $\mathcal{F}_h$ is the linear eigenvalue solver from {\sc Gene}.
The uncertain inputs in both test cases are modelled as independent uniform random variables, i.e., $\boldsymbol{\pi}$ is a multivariate uniform density, with left and right bounds stemming from expert opinion supported by experimental measurements.
All linear eigenvalue simulations are performed using $32$ cores on two Intel Xeon E5-2697 nodes from the CoolMUC-2 cluster \footnote{\texttt{https://www.lrz.de/services/compute/linux-cluster/}} at the Leibniz Supercomputer Center.
The runtime of one simulation varies roughly between five and $90$ minutes, depending on the used setup.

In Section \ref{subsec:cbc_test_case}, we consider a modified gyrokinetic benchmark to obtain initial insights into the behaviour of the proposed approach. 
To test the usefulness of the proposed approach in real-world plasma microinstability analysis problems, we consider in Section \ref{subsec:AUG} a particular discharge of the ASDEX Upgrade experiment.
For this second test case, the analysis is performed step-wisely -- first, only three uncertain input parameters are considered in Section \ref{subsubsec:AUG_3D}, whereas 12 stochastic parameters are taken into account in Section \ref{subsubsec:AUG_12D}.

\subsection{Modified Cyclone Base Case} \label{subsec:cbc_test_case}

The first test case is based on the so-called Cyclone Base Case (CBC) of \cite{Di00}. 
CBC has been selected since it is a popular benchmark in the gyrokinetic community and since its parameters are known to display a significant sensitivity to changes of the temperature and density gradients, which calls for a UQ approach.
The original CBC benchmarks have been restricted to one gyrokinetic ion (\emph{i}) species, assuming an adiabatic electron response and thus only electrostatic fluctuations.
However, we modify the setting to allow for more choices of stochastic parameters to better resemble realistic applications.

The extensions and deviations from the parameters from \cite{Di00} and the additionally educated assumptions for the uncertainties are summarized in Table \ref{tab:cbc_8D}. 
The electrons (\emph{e}) are treated fully gyrokinetically such that their logarithmic temperature gradients, $-L_{s}\partial_x \ln T_e$, and density gradients, $-L_{s}\partial_x\ln n_e$, need to be considered as well. 
While the former is taken in the same range as the ion temperature gradient, $-L_{s}\partial_x \ln T_i$, but can be varied independently, the logarithmic density gradient is forced to the exact value of the ion counterpart due to the quasi-neutrality constraint in plasma physics. 
The logarithmic density gradient also fixes the density ratio to $1$ while the temperature ratio, $T_i/T_e$, can be varied. Adding an electron species furthermore allows to consider electromagnetic effects in the gyrokinetic simulations. 
Their strength is determined by the kinetic-to-magnetic pressure ratio, $\beta$, which is therefore taken as another stochastic input. 
Another important parameter which is often avoided in benchmarks due to rather different implementations is the collision operator. 
Here, we employ a linearized Landau-Boltzmann collision operator and vary the corresponding normalized collision frequency, $\nu_c$, as listed in Table \ref{tab:cbc_8D}. 
Uncertainties in the circular magnetic ($\hat{s}-\alpha$) equilibrium can be considered by attributing lower and upper limits to the safety factor, $q$, the ratio of toroidal turns of a magnetic field line per poloidal turn, and its normalized radial derivative, the magnetic shear, $\hat{s} = r/q \partial q/\partial r$, where $r$ is the radial coordinate labeling flux surfaces.
To summarize, we consider eight uncertain input parameters: the first six characterize the underlying particle species, ions and electrons, whereas the last two parameters are associated to the magnetic geometry.
\begin{table}
\centering
\begin{tabular}{|c|c|c|cc|}
\hline
$\boldsymbol{\theta}$ & parameter name & symbol & left bound  & right bound \\ 
\hline
$\theta_1$ & plasma beta & $\beta$ & $0.598 \times 10^{-3}$ & $0.731 \times 10^{-3}$ \\
$\theta_2$ & collision frequency & $\nu_c$ & $0.238 \times 10^{-2}$ & $0.322 \times 10^{-2}$ \\
$\theta_3$ & $i$ log temperature gradient & $-L_{s}\partial_x \ln T_i$ & $7.500$ & $12.500$ \\
$\theta_4$ & $e$ log temperature gradient & $-L_{s}\partial_x \ln T_e$ & $7.500$ & $12.500$ \\
$\theta_5$ & temperature ratio & $T_i/T_e$ & $0.950$ &  $1.050$ \\
$\theta_6$ & $i$/$e$ log density gradient & $-L_{s}\partial_x\ln n$ & $1.665$ & $2.775$ \\
\hline
$\theta_7$ & magnetic shear & $\hat{s} = \frac{r}{q} \frac{\partial q}{\partial r}$ & $0.716$ & $0.875$ \\
$\theta_8$ & safety factor & $q$ & $1.330$ & $1.470$ \\
\hline
\end{tabular}
\caption{The eight stochastic parameters in the modified CBC test case.
The first six parameters characterize the two particle species, ions and electrons, whereas the last two inputs characterize the magnetic geometry.
The temperature gradient is varied per species while the density gradients of the two particles ar always equal to each other due to the quasi-neutrality condition in plasma physics. \label{tab:cbc_8D}}
\end{table}

The bounds of the eight uniformly distributed stochastic parameters are found in the last two columns in Table \ref{tab:cbc_8D} are symmetric around a nominal value employed in deterministic simulations.
The output of interest is the growth rate, $\gamma[c_{\mathrm{s}}/L_{\mathrm{s}}]$, of the dominant eigenmode, where $c_{\mathrm{s}}=\sqrt{T_e/m_i}$ is the \emph{ion sound speed} and $L_{\mathrm{s}}$ is the \emph{characteristic length}, which we compute with six digits of precision.
For a more in-depth overview of the influence of the eight stochastic parameters, we perform uncertainty propagation for multiple \emph{normalized perpendicular wavenumbers}, $k_y \rho_s$:
\begin{equation} \label{eq:kymin_CBC}
\begin{split}
k_y \rho_s &= 0.1, 0.2, 0.3, 0.4, 0.5, 0.6, 0.7, 0.8, 0.9.
\end{split}
\end{equation}
We treat $k_y \rho_s$ as a deterministic free-parameter and for each of its values in \eqref{eq:kymin_CBC}, we propagate the uncertainty in the eight parameters listed in Table \ref{tab:cbc_8D} using Algorithm \ref{algo:main_algo}.
Note that since simulations corresponding to different $k_y \rho_s$ are independent of each other, they can be performed embarrassingly parallel, thus adding a second layer of parallelism.

\begin{figure}[htbp]
  \centering
  \includegraphics[width=0.8\textwidth]{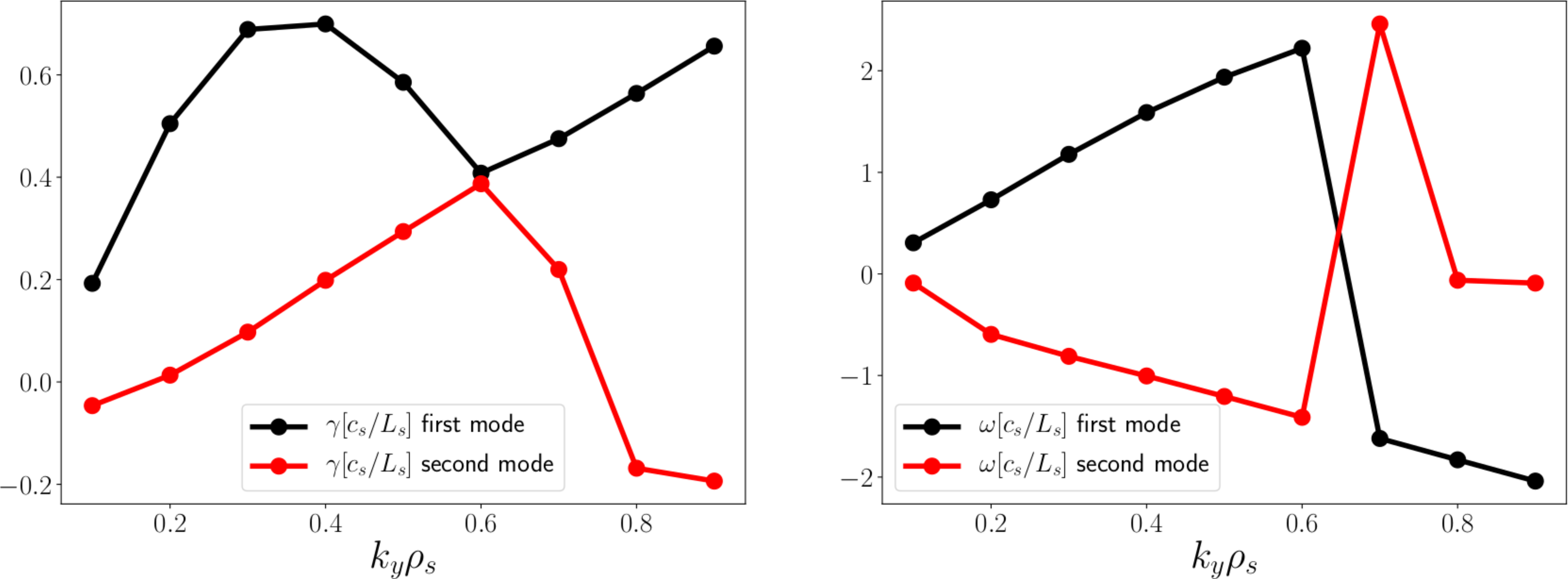}
  \caption{Growth rate (left) and frequency spectra (right) for the CBC test case of the dominant and first subdominant mode, obtained using the nominal values, i.e. the expectation of the uncertain parameters listed in Table \ref{tab:cbc_8D}.}
  \label{fig:CBC_det_plot}
\end{figure}

Prior to uncertainty propagation, a deterministic simulation using the nominal values of the eight parameters from Table \ref{tab:cbc_8D} is performed in a first step in order to obtain more insights into the underlying microinstabilities. 
We compute the linear growth rates and real frequencies, $\omega[c_{\mathrm{s}}/L_{\mathrm{s}}]$, of the dominant and first subdominant  eigenmodes for all for all $k_y \rho_s$ in \eqref{eq:kymin_CBC}.
We depict the results in Figure \ref{fig:CBC_det_plot}.
Around $k_y \rho_s = 0.6$, a change in the frequency sign is clearly visible; the positive frequency indicates a microinstability propagating in the ion-diamagnetic drift direction, whereas negative frequency is associated to a mode with (opposite) electron-diamagnetic drift direction.
For the CBC benchmark, it is well known that this mode transition is from ITG to a trapped-electron-mode(TEM)/ETG hybrid mode, see, e.g., \cite{Goe16} and the references therein.
Given the steep increase of the latter at large wavenumbers, the transition is marked by a very sharp growth rate gradient at $k_y \rho_s = 0.6$. 
Since in this work we employ sparse grid approximations formulated in terms of global polynomial basis functions, these approximations were ill-conditioned at $k_y \rho_s = 0.6$.
Therefore, we discard $k_y \rho_s = 0.6$ and present in what follows results at the remaining values of the normalized perpendicular wavenumber in \eqref{eq:kymin_CBC}.

We consider both interpolation and PSP to construct adaptive sparse grid surrogates for the $8$D stochastic model.
In addition, we compare our approach with the standard adaptive technique summarized in Section \ref{subsec:std_da}, which we use to obtain reference results.
In our approach, we prescribe $\boldsymbol{\tau}_{\mathrm{op}} = 10^{-6} \cdot \boldsymbol{1}_9$ and $\ell_{\mathrm{max}, \mathrm{op}}=20$, whereas in the standard adaptive approach we employ $tol_{\mathrm{op}} = 10^{-3}$ and $\ell_{\mathrm{max}, \mathrm{op}}=20$.
After computing the pseudo-spectral coefficients, the expectation, standard deviation and total Sobol' indices(recall Section \ref{sec:postproc}) are evaluated in the postprocessing for each considered $k_y \rho_s$.

\begin{figure}[htbp]
  \centering
  \includegraphics[width=0.7\textwidth]{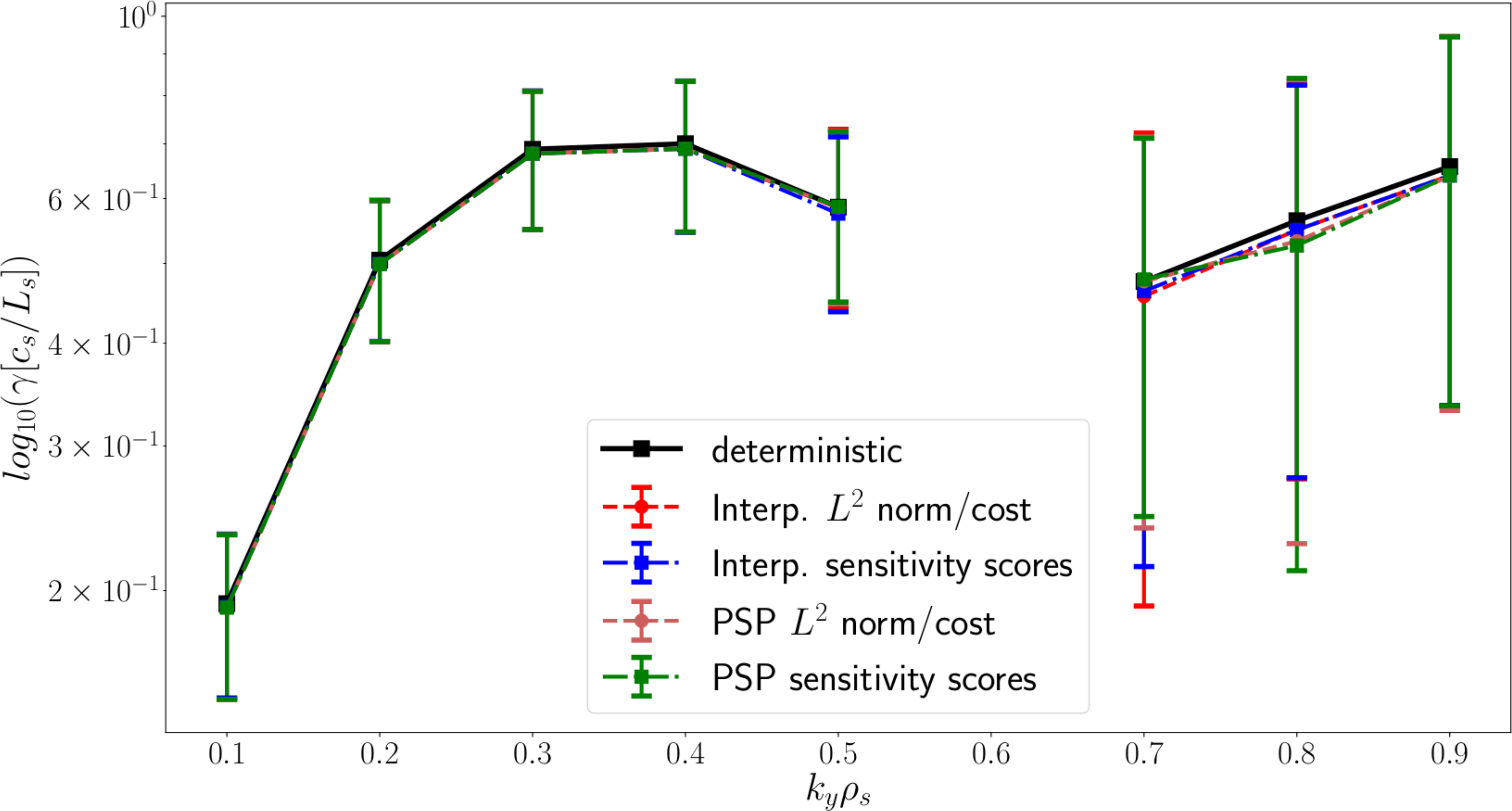}
  \caption{Expected value and one standard deviation as well as the deterministic growth rates for the CBC test case with eight uncertain parameters.}
  \label{fig:CBC_8D_error_plot}
\end{figure}

Figure \ref{fig:CBC_8D_error_plot} illustrates the expected value and one standard deviation of $\gamma[c_{\mathrm{s}}/L_{\mathrm{s}}]$ for the different adaptive surrogates as well as the deterministic growth rate results, for all considered normalized perpendicular wavenumbers.
On the one hand, a good agreement between all adaptive sparse approximation approaches can be stated.
This demonstrates that for identical setups, interpolation and PSP perform equally well.
In addition, our sensitivity scores approach is as accurate as the standard adaptive approach.
On the other hand, the deterministic results overlap almost perfectly with the expectation of the stochastic results, which is no surprise, given the relative simplicity of this test case.
Therefore, the most likely value of the growth rate yielded by uncertainty propagation is similar to the deterministic result. 
The novelty of using uncertainty propagation is that we also obtain
uncertainty bars which represent a quantitative measure of the uncertainty associated to the expected value of each stochastic input. 
Based on these results, the ITG mode appears to be robust w.r.t.~the uncertainty in the eight stochastic parameters while the onset of the TEM/ETG-branch could be quite different given the much larger uncertainty bars. 
Already these results can be very interesting for physicists trying to compare and predict microturbulence in plasmas. However, the analysis can be taken even further.
To quantitatively understand the total contribution of each stochastic input to this uncertainty, we analyse in a next step the total Sobol' indices for global sensitivity analysis.

\begin{figure}[htbp]
  \centering
  \includegraphics[width=1.0\textwidth]{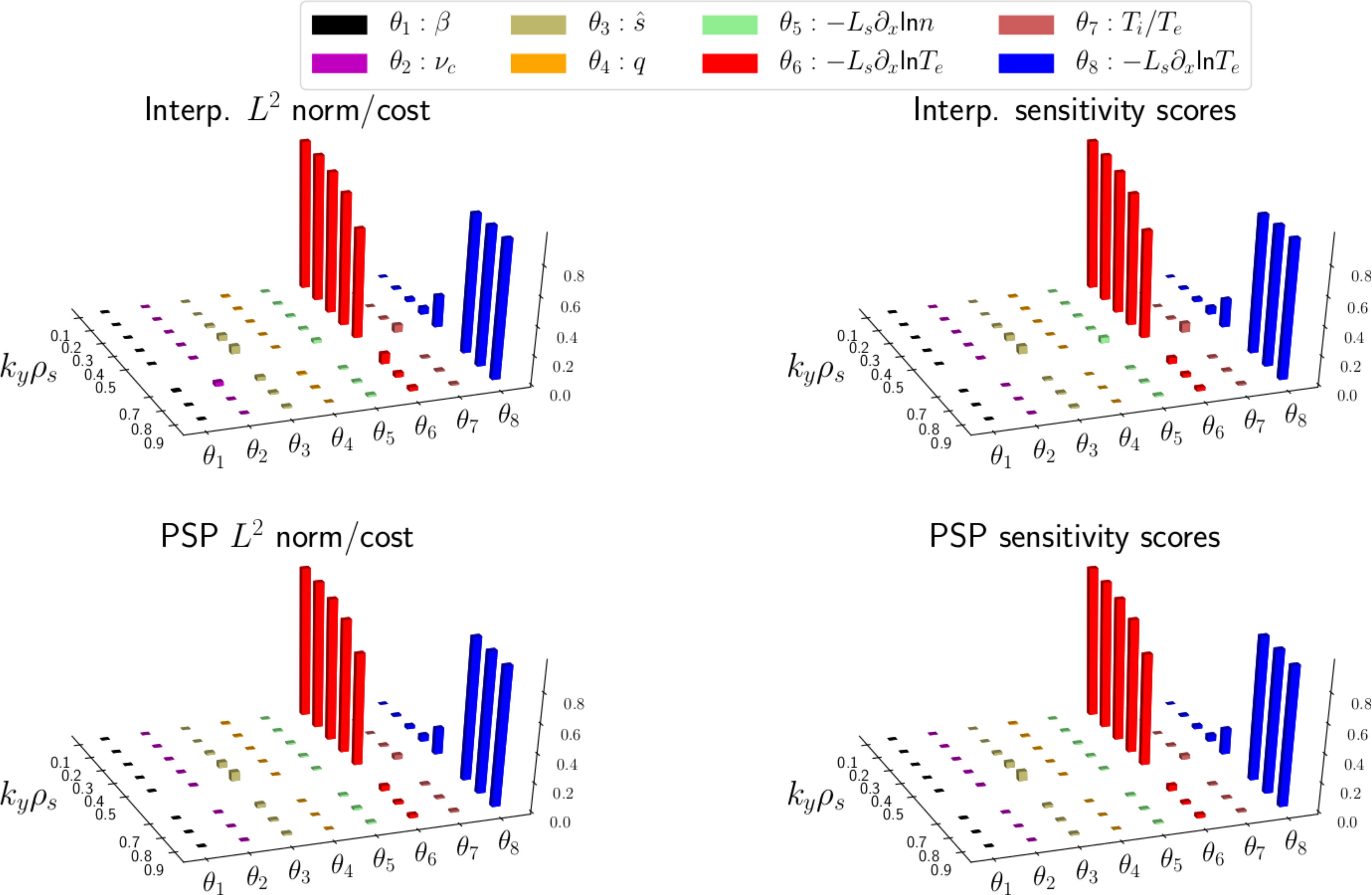}
  \caption{Total Sobol' indices for global sensitivity analysis via adaptive sparse grid interpolation (top two subfigures) and PSP (bottom two subfigures) based on the standard adaptive approach (left plots) and our proposed approach (right plots) for the CBC test case using $k_y \rho_s \in \{0.1, 0.2, 0.3, 0.4, 0.5, 0.7, 0.8, 0.9\}$.}
  \label{fig:CBC_8D_TSI}
\end{figure}
We show the total Sobol' indices in Figure \ref{fig:CBC_8D_TSI}.
The top two subfigures contain the results using adaptive interpolation, whereas in the bottom subfigures total Sobol' indices obtained by adaptive PSP are presented.
In addition, the left figures correspond to the standard adaptive approach, whereas the right plots depict the results corresponding to the proposed sensitivity scores approach.
The similarity of all results indicates again that interpolation and PSP perform similarly, and that our proposed adaptive approach has a similar accuracy as the standard adaptive method.
Moreover, we observe that in all plots two stochastic parameters show the largest total Sobol' indices, the logarithmic ion and electron temperature gradients. 
The other inputs have negligible contributions.
This indicates that although we considered a total of eight stochastic inputs, only two of them are important for uncertainty propagation in the considered $k_y \rho_s$ domain. Such findings are very important for the physics community. The other stochastic parameters are very well known to have an impact on the modes themselves but here apparently not within the assumed uncertainty bars. 
This could, for instance, motivate to restrict the much more costly nonlinear studies to the two temperature gradients if a compromise between computationally resources and sensitivity studies needs to be found.
In addition, the two temperature gradients have complementary total Sobol' indices: the ion temperature gradient dominates whereas the electron temperature gradient has a very small total Sobol' index for $k_y \rho_s \leq 0.5$, and the other way around for $k_y \rho_s \geq 0.7$.
In other words, except for $k_y \rho_s = 0.5$, where the Sobol' indices for both temperature gradients are non-negligible, the intrinsic stochastic dimensionality is one.
The observed behaviour of the total Sobol' indices is consistent with what is known in the deterministic case: for $k_y \rho_s \leq 0.5$, the microinstability is driven by an ITG mode, whereas when $k_y \rho_s \geq 0.7$, it is driven by a TEM/ETG mode.
Furthermore, at each $k_y \rho_s$ the sum of the total Sobol' indices is close to $1.0$, i.e., the stochastic model can be well approximated by a linear model. 
Such information can be extremely helpful in constructing reduced, e.g., quasi-linear models in the plasma physics community.

\begin{figure}[htbp]
  \centering
  \includegraphics[width=0.7\textwidth]{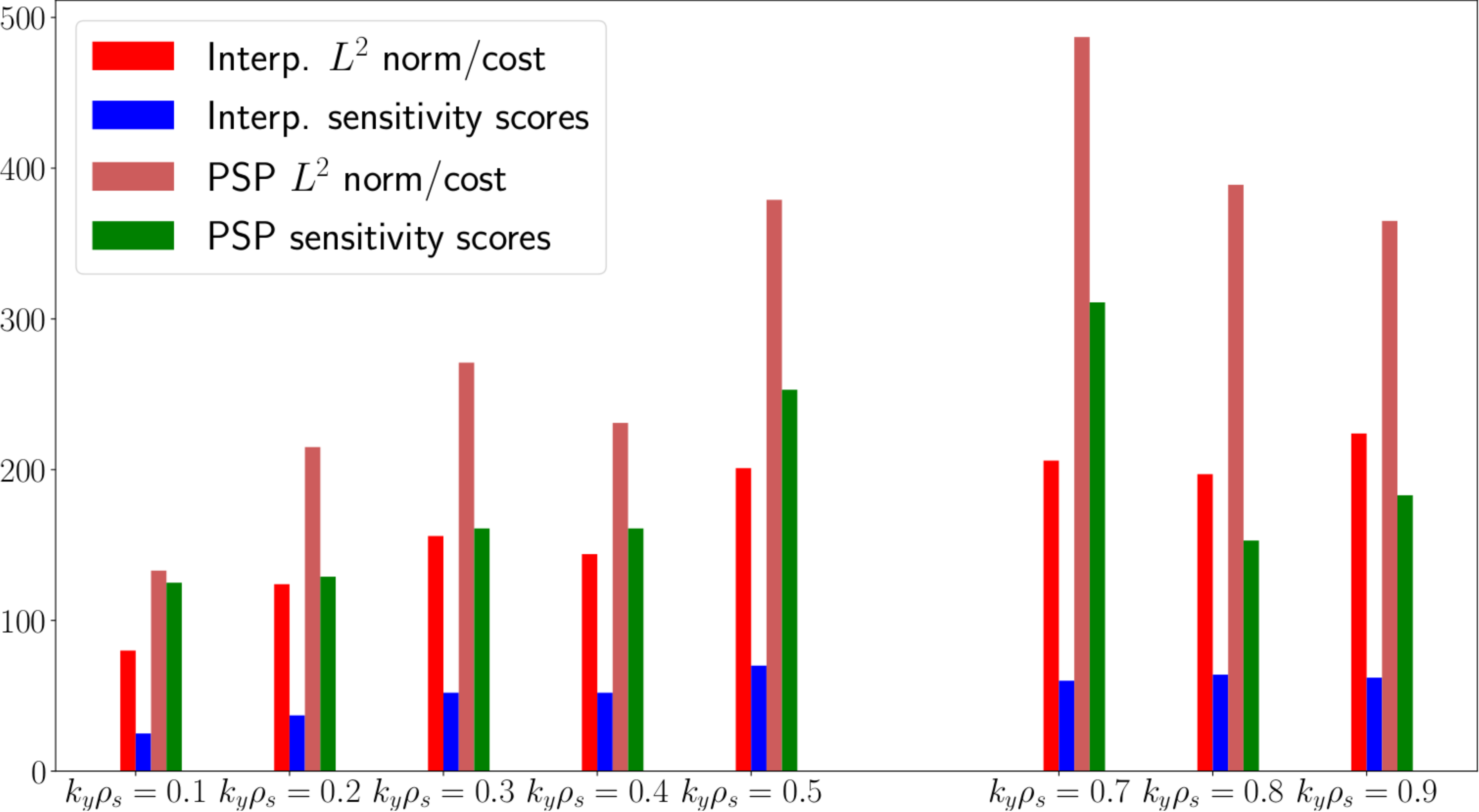}
  \caption{Total number of evaluations needed to construct the sparse interpolation and PSP surrogates using our proposed approach based on sensitivity scores vs. the standard adaptive approach.}
  \label{fig:CBC_8D_cost}
\end{figure}

Finally, the cost in terms of total number of simulations for all four employed adaptive approaches is summarized in Figure \ref{fig:CBC_8D_cost}.
First, we observe that the most expensive problem required just $473$ simulations in total; this was mainly possible because the stochastic inputs are anisotropically coupled and the intrinsic stochastic dimensionality is significantly lower than eight.
To put this in perspective, a standard parameter scan using $10$ points in each direction would require a total of $10^8$ simulations.
In addition, as expected, for either standard or sensitivity scores adaptivity, PSP is more expensive than interpolation (see Remark \ref{re:remark_spectral_interp_basis}); even the adaptive PSP based on sensitivity scores is generally more expensive than standard adaptive interpolation.
Furthermore by far the cheapest approach is the adaptive interpolation using our proposed sensitivity scores approach.
For example, at $k_y \rho_s = 0.7$, it is $3.4$ times cheaper than interpolation using standard adaptivity, $8.1$ times cheaper that adaptive PSP based on the standard method, and $5.2$ times cheaper than PSP with adaptivity based on sensitivity scores.
As a conclusion, although PSP and interpolation have very similar accuracies, the significantly lower cost of interpolation makes it our method of choice in the following realistic and computationally more expensive test case.

\subsection{Realistic test case} 
\label{subsec:AUG}

While the CBC benchmark is a popular parameter set, corresponding nonlinear simulations are found to dramatically overestimate the transport levels obtained in the experiment that inspired this exercise. 
This can to some degree be explained by idealizations in the choice of parameters and by missing physics such as external shear flows. 
The level of realism shall therefore be increased by considering a particular {\sc Gene} validation study from \cite{Fr18}, performed for the ASDEX Upgrade tokamak. 
Therein, nonlinear simulation results have been confronted to experimentally determined ion and electron heat fluxes as well as various turbulence observables such as electron temperature fluctuation levels, radial correlation functions and cross phases with electron density fluctuations which became accessible via a new diagnostic. 

Linear simulations furthermore revealed that the parameters associated to the discharge of interest are very close to the dominant mode transitions and therefore represent a challenging set which could (a) be influenced by further stochastic parameters and (b) turn out to be subject to cross interactions among these parameters. 
We took these concerns as motivation to perform a two-step UQ analysis. 

\subsubsection{Realistic test case with three stochastic parameters}
\label{subsubsec:AUG_3D}

First, we perform uncertainty propagation using three stochastic parameters as listed in Table \ref{tab:aug_3D}.
The three stochastic parameters are the logarithmic density and temperature gradients for ions and electrons.
As for the CBC test case, the left and right bounds  are symmetric around a nominal value used in deterministic simulations (see the last two columns in Table \ref{tab:aug_3D}).
However, for this test case the nominal values stem from experimental measurements. 
\begin{table}[htbp]
\centering
\begin{tabular}{|c|c|c|cc|} 
\hline
$\boldsymbol{\theta}$ & parameter name & symbol & left bound  & right bound \\ 
\hline
$\theta_1$ & $i$/$e$ log density gradient& $-L_{s} \partial_x \ln n$  & $1.156$ & $1.927$ \\
$\theta_2$ & $i$ log temperature gradient & $-L_{s}\partial_x \ln T_i$ & $2.096$ & $3.494$ \\
$\theta_3$ & $e$ log temperature gradient & $-L_{s}\partial_x \ln T_e$ & $4.040$ & $6.733$ \\
\hline
\end{tabular} 
\caption{Summary of the three stochastic parameters considered for the ASDEX Upgrade test case with ranges estimated from experimental measurements.
\label{tab:aug_3D}}
\end{table}

We perform uncertainty propagation for multiple values of $k_y \rho_s$.
Because this test case is more representative for real-world problems, wave-numbers up to the hyper-fine electron gyroradius scales are considered:
\begin{equation} \label{eq:kymin_AUG}
\begin{split}
k_y \rho_s &= 0.2, 0.3, 0.4, 0.5, 0.6, 0.7, 0.9, 1.0, \\
& 2.0, 3.0, 4.0, 5.0, 7.5, 10.0, 12.5, 15.0, 17.5, 20.0, 22.5, 25.0, 27.5, 30.0.
\end{split}
\end{equation}
As in the CBC test case, we treat $k_y \rho_s$ as a deterministic free-parameter and perform uncertainty propagation using Algorithm \ref{algo:main_algo} for each value in \eqref{eq:kymin_AUG}.

\begin{figure}[htbp]
  \centering
  \includegraphics[width=0.8\textwidth]{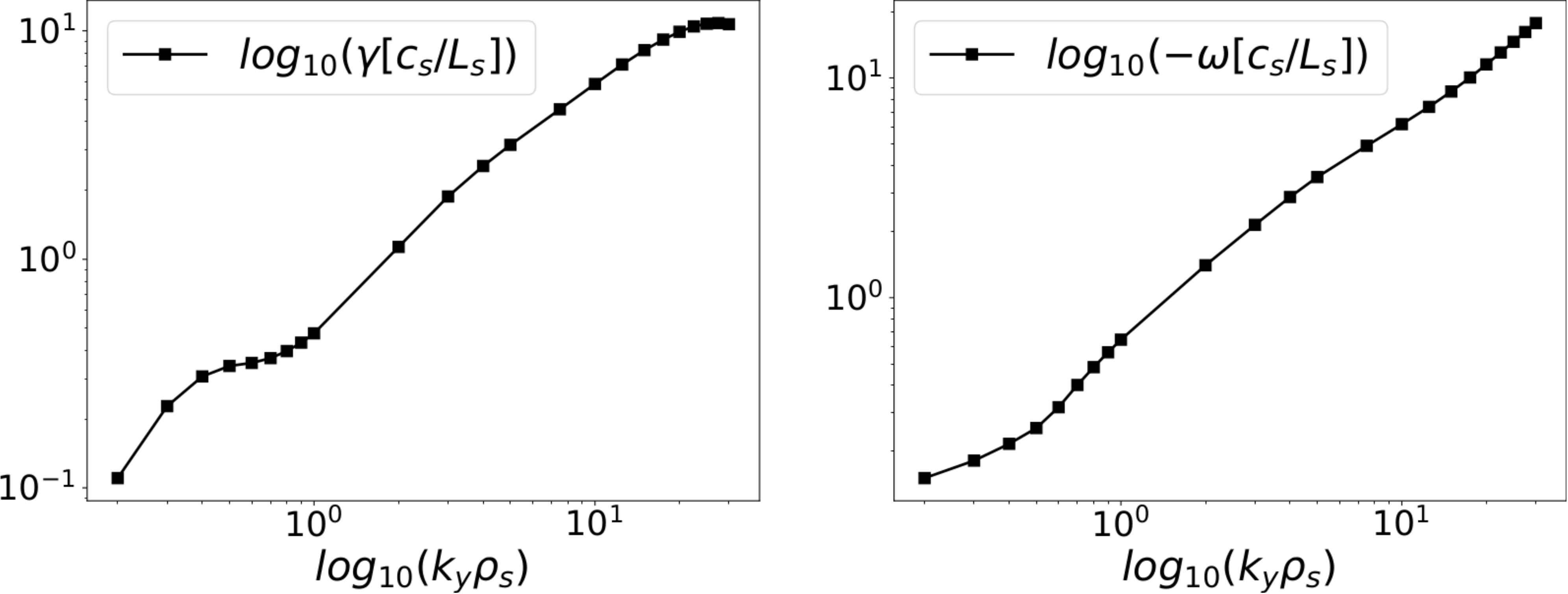}
  \caption{Growth rate (left) and frequency spectra (right) for the ASDEX Upgrade test case obtained using the nominal values of the input parameters. Note that for this test case, these values stem from experimental measurements.}
  \label{fig:ASDEX_det_plot}
\end{figure}

To obtain some first insights into the underlying microinstabilities, deterministic runs using the nominal values of all input parameters, which stem from experimental data, are performed to compute the growth rate, $\gamma[c_{\mathrm{s}}/L_{\mathrm{s}}]$, and frequency spectra, $\omega[c_{\mathrm{s}}/L_{\mathrm{s}}]$, of the dominant eigenmode.
Given the large wave number and amplitude range, the results are depicted in a logarithmic scale in Figure \ref{fig:ASDEX_det_plot} (note the negative sign of the frequency).
Both curves are smooth and monotonous, suggesting that the dominant mode does neither change its (electron diamagnetic) drift direction nor its character in a dramatic way for the parameters at hand. 
The slightly different slopes in frequency, i.e., dispersion relations, and the first hump in growth rate imply that nevertheless two or more different microinstabilities such as pure TEM and TEM/ETG-hybrids are excited.

In Section \ref{subsec:cbc_test_case}, we concluded that adaptive sparse grid interpolation is our method of choice for this realistic test case.
Hence, a surrogate for the 3D stochastic gyrokinetic model is constructed via adaptive interpolation using both the proposed methodology based on sensitivity scores as well as the standard adaptive technique summarized in Section \ref{subsec:std_da}.
In our proposed approach we prescribe $\boldsymbol{\tau}_{\mathrm{in}} = 10^{-6} \cdot \boldsymbol{1}_4$ and $\ell_{\mathrm{max}, \mathrm{in}}=20$, whereas in the standard adaptive approach, $tol_{\mathrm{in}} = 10^{-3}$ and $\ell_{\mathrm{max}, \mathrm{in}}=20$ are employed.
As in the CBC test case, the output of interest are the growth rate, $\gamma[c_{\mathrm{s}}/L_{\mathrm{s}}]$, of the dominant eigenmode with six digits of precision. 
Additionally, we compute its expectation, standard deviation and total Sobol' indices for each $k_y \rho_s$ in postprocessing (recall Section \ref{sec:postproc}).
Due to the larger growth rates and more simple mode structure, the runtime significantly decreased with $k_y \rho_s$.

\begin{figure}[htbp]
  \centering
  \includegraphics[width=0.7\textwidth]{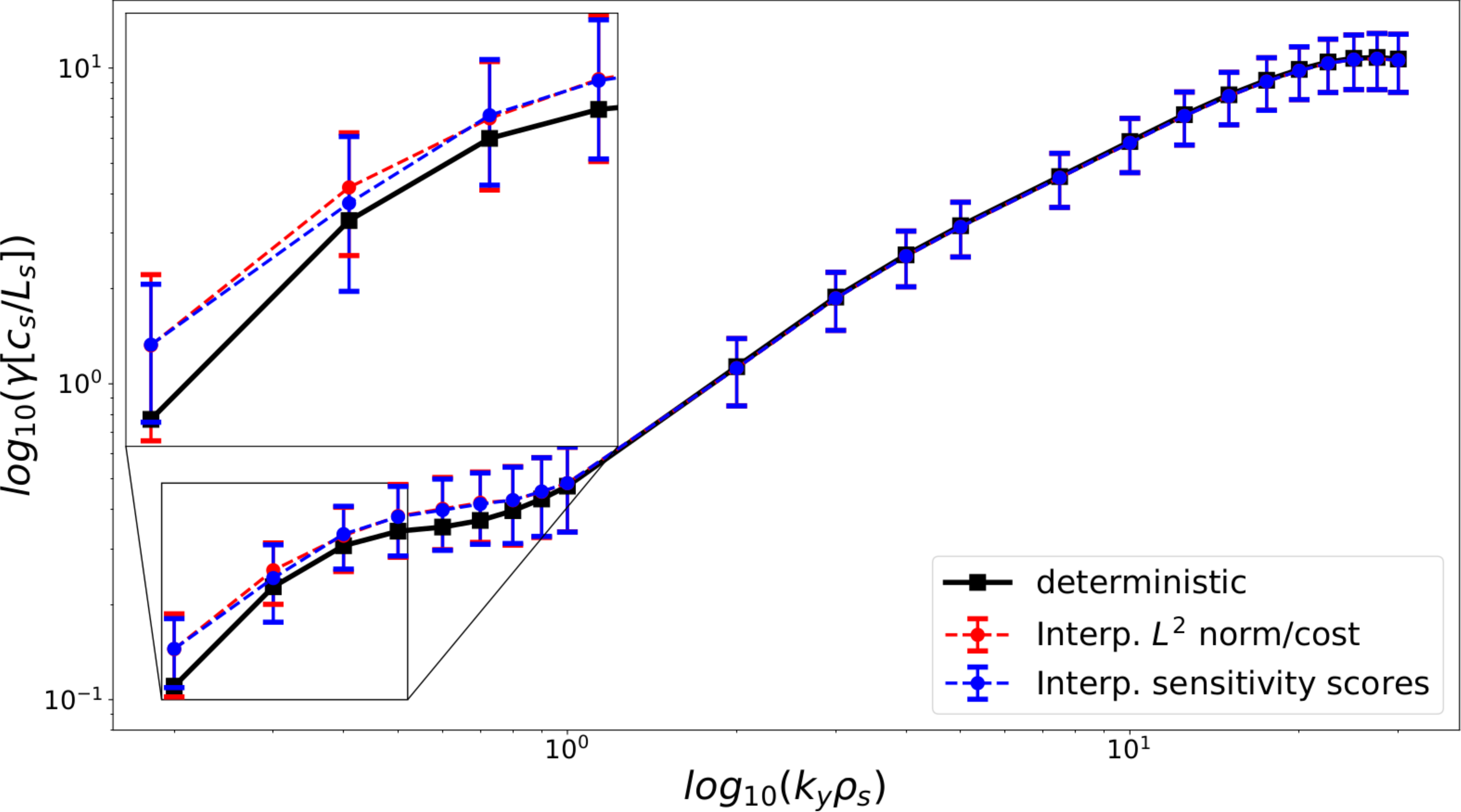}
  \caption{Expected value and one standard deviation as well as the deterministic growth rates for the ASDEX Upgrade test case with three uncertain parameters and $k_y \rho_s$ as in \eqref{eq:kymin_AUG}.}
  \label{fig:AUG_3D_error_plot}
\end{figure}

The resulting expectation and one standard deviation as well as the deterministic growth rates are displayed in Figure \ref{fig:AUG_3D_error_plot}.
An overlap between the results obtained with the two adaptive interpolation approaches is observed, showing that our proposed approach is as accurate as the standard adaptive method for this test case as well.
In addition, the deterministic result is similar to the expectation of the stochastic simulation; the absolute difference varies uniformly roughly between $0.01$ and $0.1$.
However, these differences are more significant for $k_y \rho_s \leq 1.0$ as in this region they represent a larger fraction of the growth rate's magnitude. 
Since this wave number range is usually considered to be most important for a correct assessment of the {\em ion heat flux}, results from corresponding nonlinear simulations with nominal values should be taken with care. 
The presented analysis implies that the more likely results considering uncertainties may be somewhat different which is an important piece of information for the plasma turbulence modeling community.

\begin{figure}[htbp]
  \centering
  \includegraphics[width=1.0\textwidth]{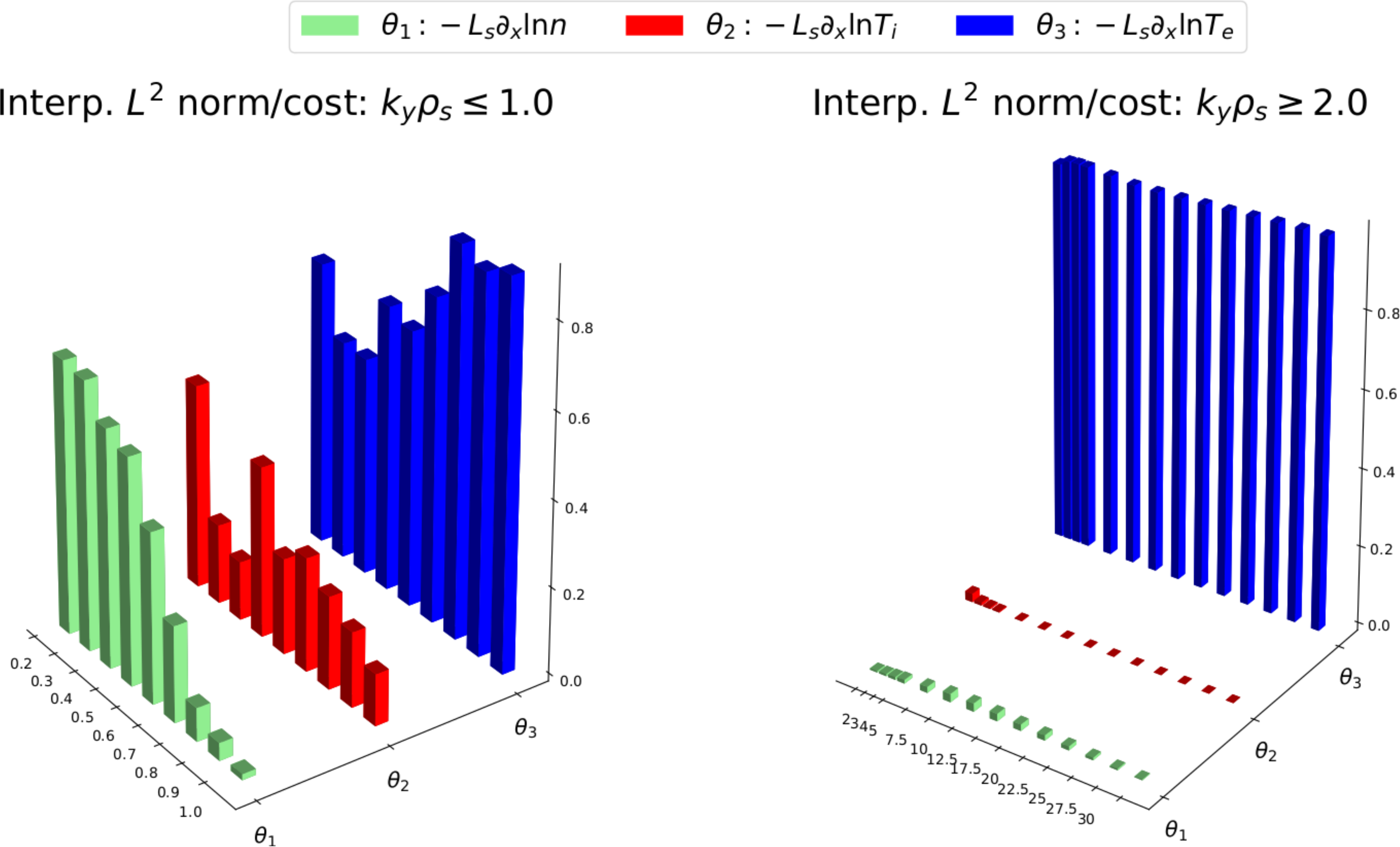}
  \includegraphics[width=1.0\textwidth]{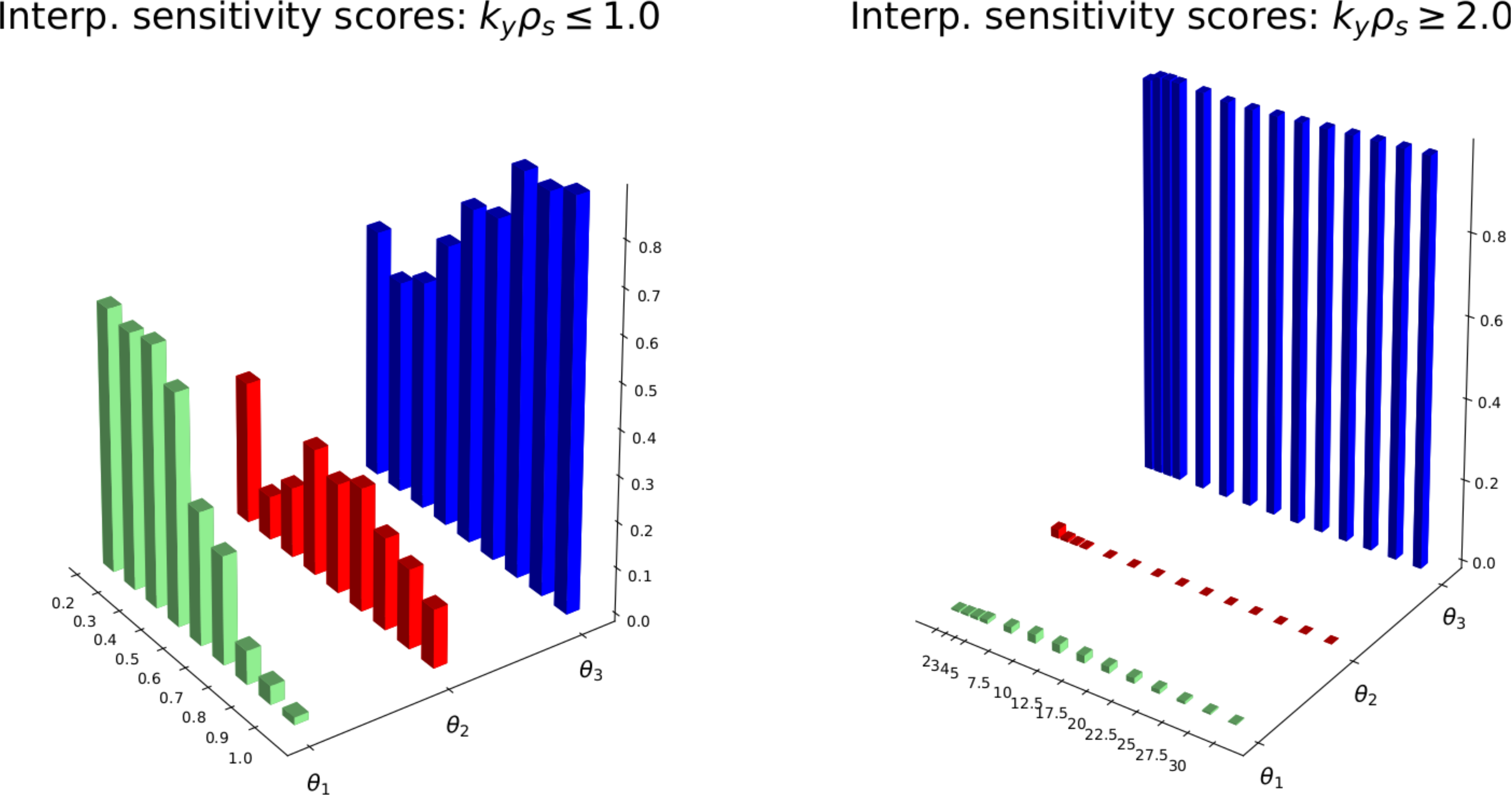}
  \caption{Total Sobol' indices for global sensitivity analysis via adaptive interpolation with the standard approach (top two subfigures) and our proposed approach (bottom two subfigures) for the ASDEX Upgrade test case with three uncertain parameters and $k_y \rho_s$ as in \eqref{eq:kymin_AUG}.}
  \label{fig:AUG_3D_TSI}
\end{figure}

For a deeper understanding of the results, the total Sobol' indices are furthermore depicted in Figure \ref{fig:AUG_3D_TSI} (top two subfigures: standard adaptive approach, bottom subplots: our proposed approach).
For a clear illustration, we split the results in two figures corresponding to $k_y \rho_s \leq 1$ and $k_y \rho_s > 1$, respectively.
Again a good agreement between our proposed approach and the standard adaptive strategy is observed. 
On the one hand, when $k_y \rho_s \leq 0.8$, all three stochastic parameters have non-negligible total Sobol' indices, whereas for $k_y \rho_s = 0.9, 1$, the two logarithmic temperature gradients are the most important, i.e., the intrinsic stochastic dimensionality is two.
For $k_y \rho_s \leq 1$, the stochastic model exhibits a nonlinear behaviour due to the non-negligible interaction between the inputs.
When $k_y \rho_s > 1$, on the other hand, the uncertainty in the logarithmic electron temperature gradient is the most important stochastic parameter.
Moreover, the sum of the total Sobol' indices is close to $1$, which shows that the stochastic model can be well approximated by a linear model.
In other words, when $k_y \rho_s > 1$ the intrinsic stochastic dimension is one.
Figure \ref{fig:AUG_3D_TSI} also provides information on the underlying microinstabilities.
For $k_y \rho_s > 1$, we clearly have an ETG mode: the electron temperature gradient is most important parameter.
For $k_y \rho_s \leq 1$, on the other hand, the electron temperature gradient is more important than the ion temperature gradient and the contribution of the density gradient decreases as $k_y \rho_s$ increases. Hence, we have a mixture of TEM/ETG mode for $k_y \rho_s \leq 1$ which is, however, also affected by the ion temperature gradient and may be in competition with subdominant ITG modes.

\begin{figure}[htbp]
  \centering
  \includegraphics[width=0.7\textwidth]{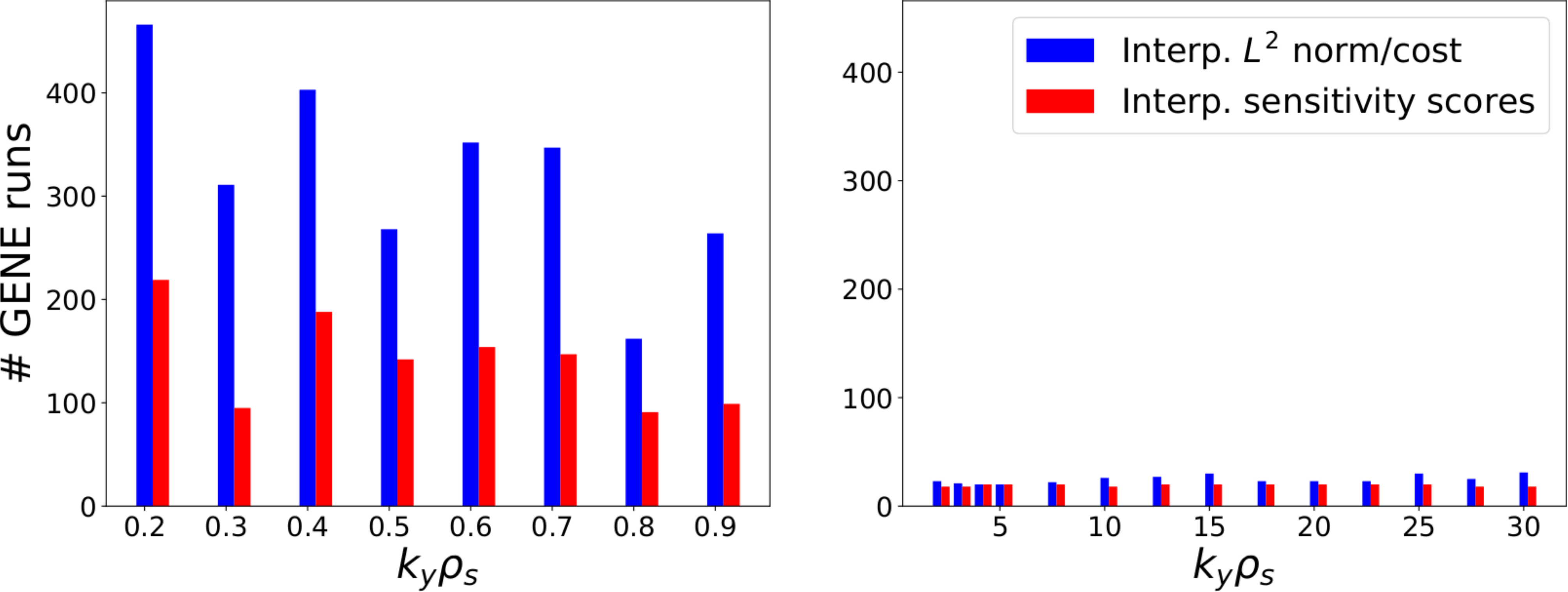}
  \caption{Total number of {\sc Gene} evaluations needed to construct the sparse interpolation surrogate using our proposed approach based on sensitivity scores vs. the standard adaptive approach for the ASDEX Upgrade test case with three uncertain parameters.}
  \label{fig:AUG_3D_cost}
\end{figure}

In Figure \ref{fig:AUG_3D_cost}, we visualize the cost of the two adaptive strategies in terms of total number of grid points, i.e., {\sc Gene} evaluations (left: $k_y \rho_s \leq 1$, right: $k_y \rho_s \geq 2$).
Note that the behaviour observed in the total Sobol' indices is reflected in the cost as well.
It is computationally more expensive to perform uncertainty propagation for $k_y \rho_s \leq 1$ than for $k_y \rho_s > 1$ due to the higher intrinsic stochastic dimension.
In addition, our approach requires fewer {\sc Gene} runs than the standard adaptive approach for all normalized perpendicular wavenumbers $k_y \rho_s$.
For example, at $k_y \rho_s = 3$, we need about $3.3$ times fewer {\sc Gene} evaluations that the standard approach. 
However, for $k_y \rho_s \geq 1$ the savings yielded by our approach are not very significant because both approaches are computationally cheap, requiring a small, roughly the same for all these $k_y \rho_s$, number of {\sc Gene} evaluations; this is mainly because the difference between the given stochastic dimensionality, three, is not significantly larger than the intrinsic dimensionality, one.

\begin{figure}[htbp]
  \centering
  \includegraphics[width=0.7\textwidth]{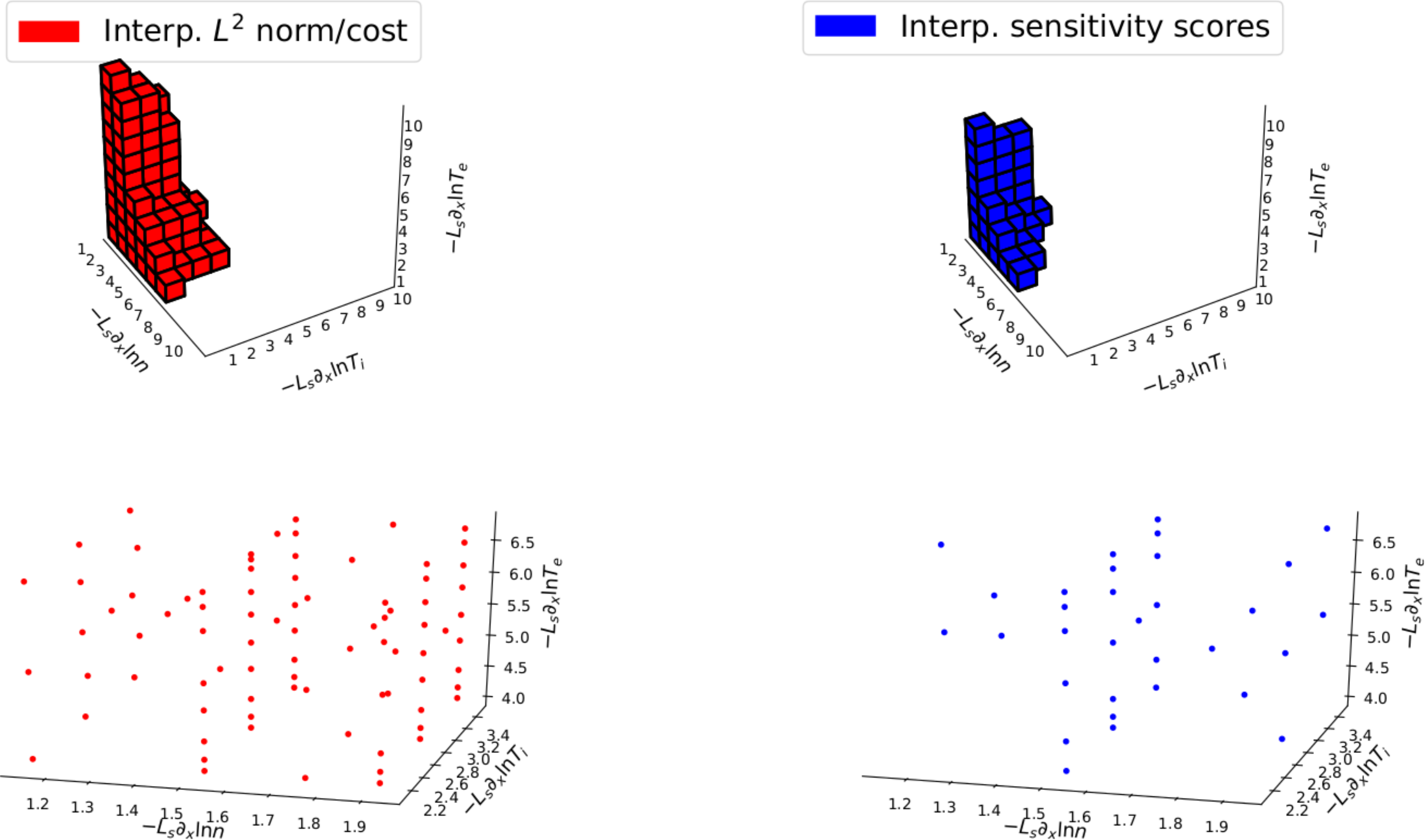}
  \caption{Multiindices and their associated sparse grids at $k_y \rho_s = 1.0$ for adaptive interpolation for the $3$D ASDEX Upgrade scenario (left: standard adaptive approach, right: our proposed approach).}
  \label{fig:AUG_3D_cost_savings}
\end{figure}

To further illustrate the computational savings due to our approach, we depict in Figure \ref{fig:AUG_3D_cost_savings} the multiindices and the associated sparse grids at $k_y \rho_s = 1.0$ (left: standard adaptive approach, right: sensitivity scores approach).
The three estimated total Sobol' indices are $\hat{S}_1^T = 0.014$, $\hat{S}_2^T = 0.120$, and  $\hat{S}_3^T = 0.900$. 
Therefore, the logarithmic electron density gradient is the most important direction, the logarithmic ion density gradient is the second, while the third important direction is the logarithmic density gradient.
The total Sobol' indices sum up to $1.033$, which indicates that the interactions between the three inputs are negligible.
The multiindex sets in the top subfigures in Figure \ref{fig:AUG_3D_cost_savings} show that both algorithms detect the third direction as the most important.
However, the standard adaptive algorithm adds many (unnecessary) interaction multiindices, while our approach better exploits the fact that one direction is significantly more important than the other two combined and that the coupling between the three inputs is negligible. 
At the end of the refinement process, the standard approach yields $83$ grid points, whereas our approach needs a total of only $32$ points to produce similar results.

\subsubsection{Realistic test case with 12 stochastic inputs} \label{subsubsec:AUG_12D}

In a second uncertainty analysis step, we enhance the previous scenario with nine additional stochastic terms shown in Table \ref{tab:aug_12D}, resulting in a total of $12$ stochastic inputs for the ASDEX Upgrade test case.
The first seven characterize the two particle species, ions and electrons, while the latter five parameters are associated to the magnetic geometry.
As for the previous test case, the uniform bounds of the stocahstic inputs are symmetric around nominal values stemming from experimental measurements.
Note that the deterministic growth rates and frequencies are the same as for the $3$D case in Section \ref{subsubsec:AUG_3D} as we have the same nominal input parameters.
\begin{table}[htbp]
\centering
\begin{tabular}{|c|c|c|cc|}
\hline
$\boldsymbol{\theta}$ & parameter name & symbol & left bound  & right bound \\ 
\hline
$\theta_1$ & plasma beta & $\beta$ & $0.488 \times 10^{-3}$ & $0.597 \times 10^{-3}$ \\
$\theta_2$ & collision frequency & $\nu_c$ & $0.641 \times 10 ^{-2}$ & $0.867 \times 10^{-2}$ \\
$\theta_3$ & $i$/$e$ log density gradient & $-L_{s}\partial_x\ln n$ & $1.156$ & $1.927$ \\
$\theta_4$ & $i$ log temperature gradient & $-L_{s}\partial_x\ln T_i$ & $2.096$ & $3.494$ \\
$\theta_5$ & temperature ratio & $T_i/T_e$ & $0.610$ &  $0.670$ \\
$\theta_6$ & $e$ log temperature gradient & $-L_{s}\partial_x\ln T_e$ & $4.040$ & $6.733$ \\
$\theta_7$ & effective ion charge & $Z_{\mathrm{eff}} = \sum_i n_i q_i^2/\sum_i n_i$ & $1.280$ & $1.920$ \\
\hline
$\theta_8$ & safety factor & $q$ & $2.170$ & $2.399$ \\
$\theta_9$ & magnetic shear & $\hat{s} = \frac{r}{q}\frac{dq}{dr}$ & $1.992$ & $2.435$ \\
$\theta_{10}$ & elongation & $k$ & $0.128$ & $0.141$ \\
$\theta_{11}$ & elongation gradient & $s_k = \frac{r}{k} \frac{\partial k}{\partial r}$ & $0.200$ & $0.250$ \\
$\theta_{12}$ & triangularity & $\delta$ & $0.710$ & $0.870$ \\
\hline
\end{tabular}
\caption{Summary of the $12$ stochastic parameters considered for the ASDEX Upgrade test case.
\label{tab:aug_12D}}
\end{table}

\begin{figure}[htbp]
  \centering
  \includegraphics[width=0.7\textwidth]{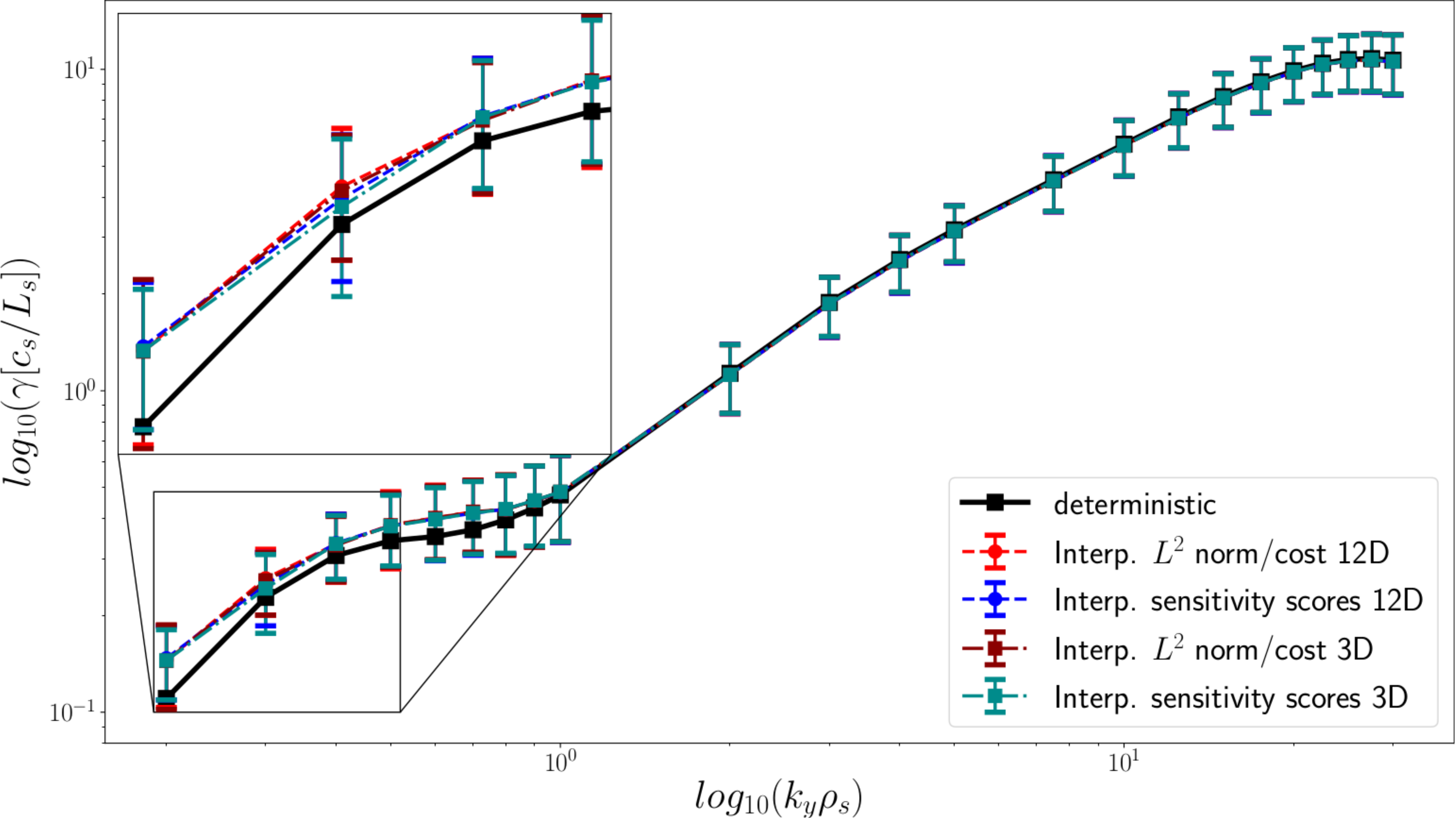}
  \caption{Expected value and one standard deviation as well as the deterministic growth rates for the ASDEX Upgrade test case with $12$ uncertain parameters and $k_y \rho_s$ as in \eqref{eq:kymin_AUG}.}
  \label{fig:AUG_12D_error}
\end{figure}

We extend the stochastic setup from the $3$D scenario for $12$ uncertain inputs: $\boldsymbol{\tau}_{\mathrm{in}} = 10^{-6} \cdot \boldsymbol{1}_{13}$, while the other tolerance and the maximum levels remain the same.
We visualize the expectation, one standard deviation and the deterministic growth rates for both the $3$D and $12$D scenarios, for comparison, in Figure \ref{fig:AUG_12D_error}.
On the one hand, we see that for this scenario too, our proposed approach yields results very similar to the standard adaptive strategy.
Moreover, the difference between the deterministic results and the error plots is quantitatively very similar to the $3$D case.
On the other hand, we observe that the expectations and standard deviations for the $12$D case overlap almost perfectly with the $3$D results. 
Therefore, we assume that the the extra $9$ uncertain parameters contribute insignificantly to the overall results which would be an important piece of information for on-going validation studies which may now mainly focus on the logarithmic gradient sensitivities in nonlinear turbulence simulations.

\begin{figure}[htbp]
  \centering
  \includegraphics[width=1.0\textwidth]{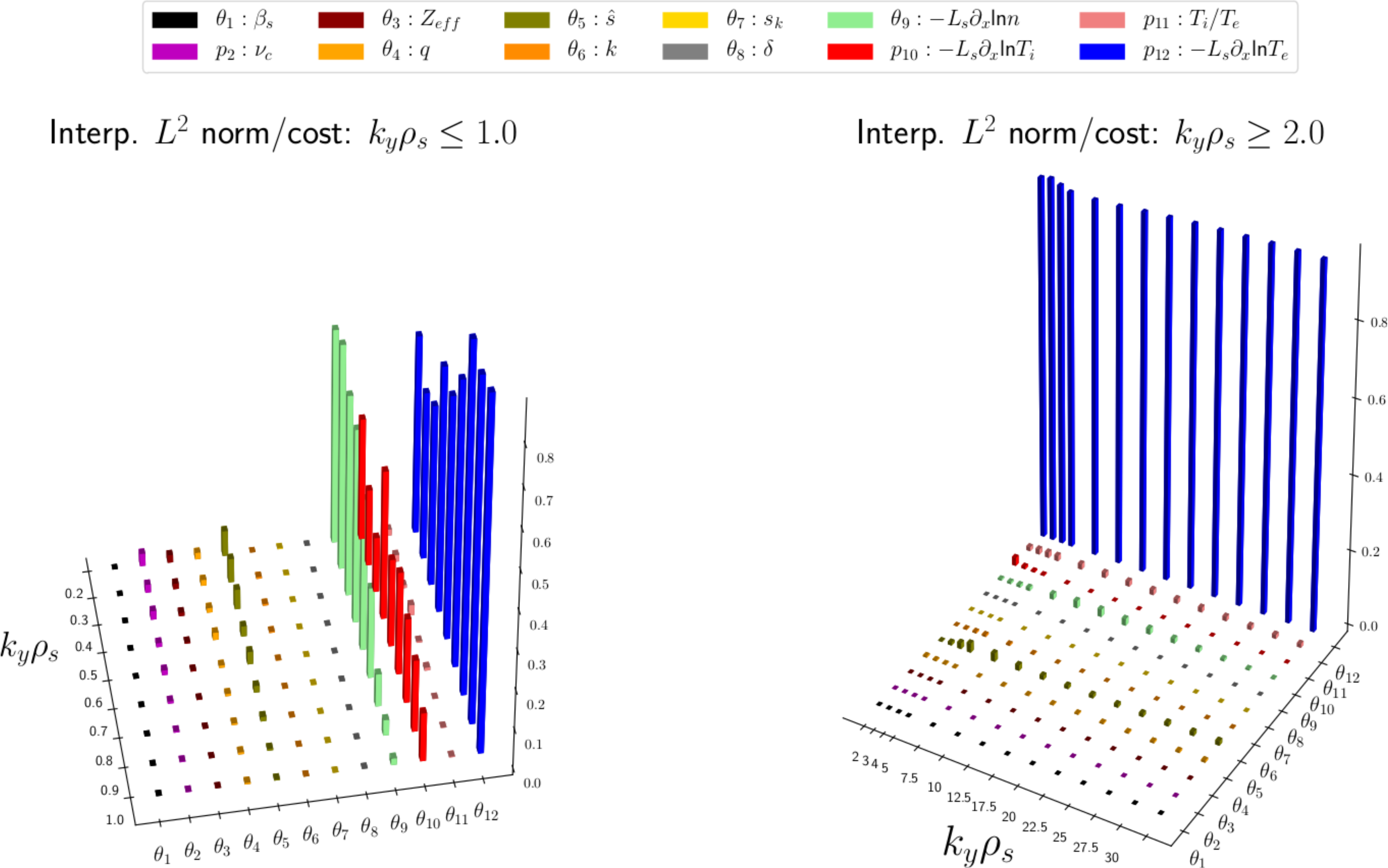}
  \includegraphics[width=1.0\textwidth]{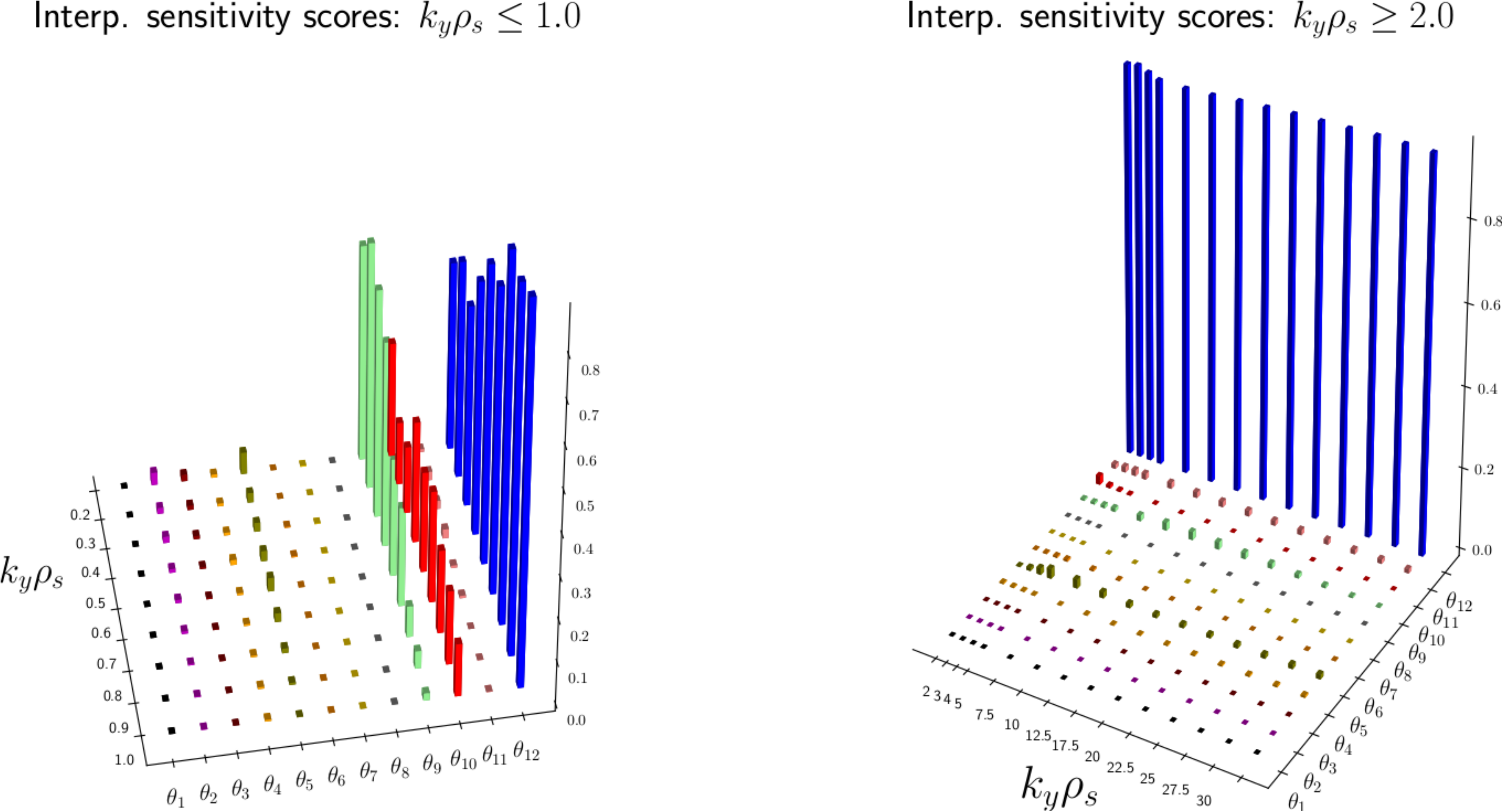}
  \caption{Total Sobol' indices for global sensitivity analysis via adaptive interpolation with the standard approach (top two subfigures) and our proposed approach (bottom two subfigures) for the ASDEX Upgrade test case with $12$ uncertain parameters and $k_y \rho_s$ as in \eqref{eq:kymin_AUG}.}
  \label{fig:AUG_12D_TSI}
\end{figure}

To ascertain the above assumption, we analyse the total Sobol' indices for sensitivity analysis in Figure \ref{fig:AUG_12D_TSI} (top: standard adaptive interpolation, bottom: adaptive interpolation with our proposed sensitivity scores approach).
Note that as for the $3$D scenario, our proposed approach produces results very similar to the standard approach.
We see that besides some small contributions due to the magnetic shear, $\hat{s}$, at $k_y \rho_s \leq 0.5$, the total Sobol' indices associated to the logarithmic temperature and density gradients have the largest values for $k_y \rho_s \leq 1.0$ while the remaining $9$ parameters have negligible Sobol' indices.
Moreover, when $k_y \rho_s \geq 2.0$, the logarithmic electron temperature gradient is the most important parameter and the other $11$ are negligible.
In addition, throughout the considered $k_y \rho_s$ domain, the total Sobol' indices of the aforementioned important parameters are qualitatively and quantitatively very similar to the indices from the $3$D scenario.
Hence, we conclude that the $9$ extra stochastic input parameters contribute insignificantly to the overall results.
In addition, the underlying microinstabilities are the same as for the $3$D scenario.

\begin{figure}[htbp]
  \centering
  \includegraphics[width=0.7\textwidth]{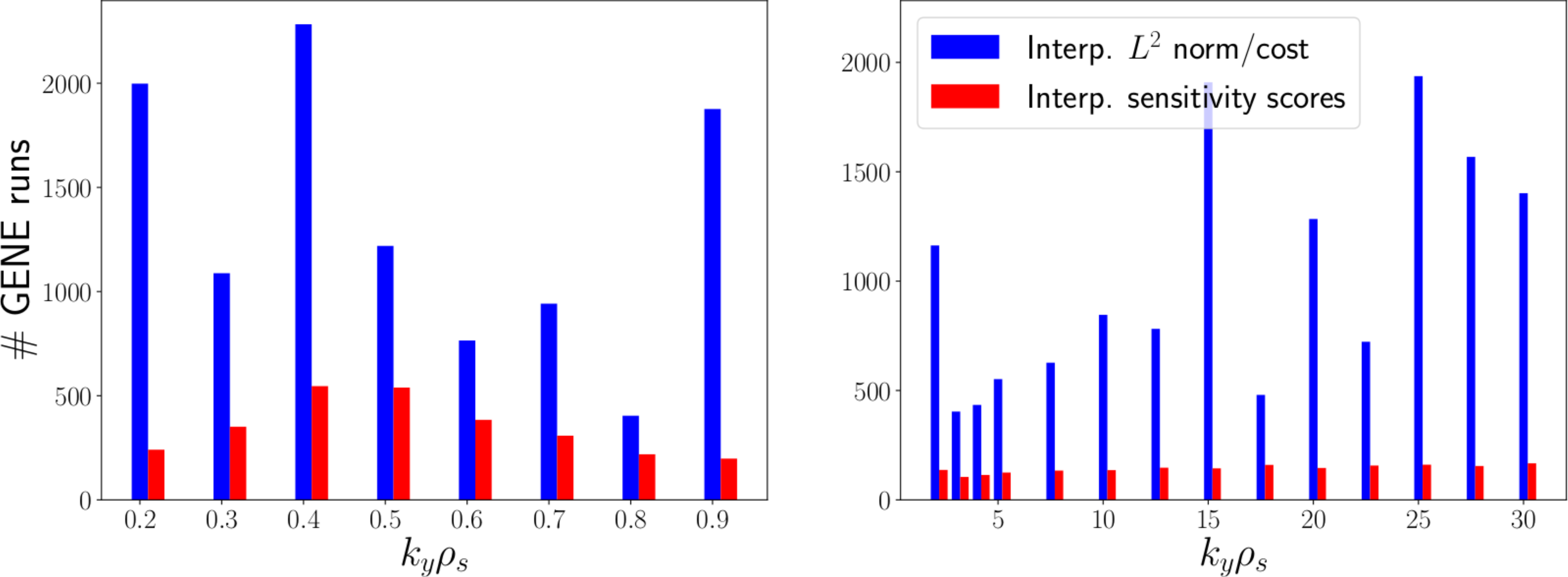}
  \caption{Total number of {\sc Gene} evaluations needed to construct the sparse interpolation surrogate using our proposed approach based on sensitivity scores vs. the standard adaptive approach.}
  \label{fig:AUG_12D_cost}
\end{figure}

In Figure \ref{fig:AUG_12D_cost}, we visualize the cost comparison between our proposed approach and the standard adaptive approach (left: $k_y \rho_s \leq 1$, right: $k_y \rho_s \geq 2$).
We observe that for all $k_y \rho_s$ the proposed method requires significantly fewer {\sc Gene} evaluations than the standard adaptive approach.
For example, at $k_y \rho_s = 15$ our approach requires $13.3$ times fewer evaluations than the standard adaptive approach.
Note, on the one hand, that for $k_y \rho_s > 1$ the cost of the sensitivity scores approach is similar for all $k_y \rho_s$, which is due to having the intrinsic stochastic dimensionality one in this region: our methodology detects that only one stochastic direction is important, hence it refines predominantly that direction.
To put the cost savings for this scenario into perspective, a full-tensor grid typically used in parameter scans constructed using $20$ points in each direction, i.e., the number of points associated to the maximum level reached by the standard adaptive approach, would require $20^{12} \approx O(10^{15})$ {\sc Gene} evaluations in total, which is computationally prohibitive.
On the other hand, the standard adaptive strategy required at most $2283$ {\sc Gene} evaluations, whereas the largest number of simulations in the proposed approach was $546$.

To conclude the two stochastic scenarios considered for the ASDEX Upgrade test case, the presented adaptive sparse approximation strategies enable uncertainty propagation and sensitivity analysis in problems in which standard approaches would be almost impossible to use.
In addition, our proposed approach proved to be as accurate as the standard adaptive approach, but at significantly reduced computational cost.
Finally, when the intrinsic stochastic dimensionality is smaller than the total number of stochastic inputs, which is usually the case in practice, our approach explores and exploits this structure to preferentially refine the important directions.

\section{Conclusions and outlook} \label{sec:conclusions}

In this work, we introduced a novel adaptive sparse grid approximation methodology for uncertainty propagation in computationally expensive real-world applications.
We leveraged Sobol' decompositions to introduce a sensitivity scoring system to drive the adaptive process.
By employing sensitivity information, we explored and exploited the anisotropic coupling of the stochastic inputs as well as the lower intrinsic stochastic dimensionality which is typical in practical applications.

The proposed approach can be formulated in terms of arbitrary approximation operators and point sets. 
Here, we considered Lagrange interpolation and pseudo-spectral projection operators which we defined using two (L)-Leja point constructions.
Although the proposed methodology is model-agnostic, a real-world problem arising in plasma turbulence research has been considered.
Besides introducing a novel adaptive methodology for quantifying uncertainty in computationally expensive problems, this work also represents, to the best of our knowledge, one of the first uncertainty propagation studies in gyrokinetic plasma microinstability analysis.

We tested and compared the proposed approach with a standard adaptive strategy in two plasma microturbulence simulation test cases.
In both test cases, the gyrokinetic code {\sc Gene} was employed to characterize the microinstabilities.
The first test case was based on the popular Cyclone Base Case benchmark of the gyrokinetic community.
Therein, we considered eight uncertain inputs characterizing both the underlying particle species, and the magnetic geometry.
The second test case was a real-world example stemming from a particular validation study for the ASDEX Upgrade tokamak device.
For this test case we carried out a two step analysis, initially considering three uncertain inputs characterizing the ions and electrons, and then $12$ stochastic parameters associated to the particle species and the magnetic geometry.
Our results showed that the proposed approach is as accurate as the standard adaptive approach while requiring a significantly reduced computational cost; for example, for the $12$D scenario, we obtained factors of up to $13.3$ fewer {\sc Gene} evaluations.
The presented adaptive sparse grid approximations enable the uncertainty propagation and sensitivity analysis in higher-dimensional plasma microturbulence problems, which would be almost impossible to tackle with standard screening approaches. It furthermore demonstrated its capability to provide deep and new insights on the parametric dependencies which will be very helpful for further comparison with experiments and construction of reduced plasma turbulence models.

For future research, we want to extend the current methodology to a multifidelity framework for uncertainty propagation (see \cite{PWG18}).
Our long-term goal with respect to plasma turbulence research is to quantify uncertainty in fully nonlinear gyrokinetic simulations.

\section*{Acknowledgements}
The first author thankfully acknowledges the support of the German Academic Exchange Service (\texttt{http://daad.de/}) and the support of the German Research Foundation and Technical University of Munich through the TUM International Graduate School of Science and Engineering (\texttt{http://igsse.tum.de/}) within the project 10.02 BAYES.
Moreover, the authors gratefully acknowledge the compute and data resources provided by the Leibniz Supercomputing Centre (\texttt{www.lrz.de}).
Part of this research was performed while the first, third and fourth authors were visiting the Institute for Pure and Applied Mathematics (IPAM), which is supported by the National Science Foundation.




\bibliographystyle{elsarticle-harv}

\bibliography{ms}

\end{document}